\documentclass[final,5p,times,twocolumn]{elsarticle}

\usepackage{amssymb}
\usepackage{amsmath} 
\usepackage{color}
\usepackage{algorithm}
\usepackage{algpseudocode}
\usepackage{float}
\usepackage{tabularx}
\usepackage{array}
\usepackage{ragged2e}
\usepackage{multirow}
\newcolumntype{Y}{>{\RaggedRight\arraybackslash}X} 

\algnewcommand{\algorithmicforeach}{\textbccf{for each}}
\algdef{SE}[FOR]{ForEach}{EndForEach}[1]
  {\algorithmicforeach\ #1\ \algorithmicdo}
  {\algorithmicend\ \algorithmicforeach}

\usepackage[switch]{lineno}

\newcommand{\mb}{\mathbf}

\renewcommand{\arraystretch}{2} 

\begin{document}

\begin{frontmatter}



\title{Topology optimization of two-fluid turbulent heat exchangers:\\A Darcy flow-based multifidelity approach}


\author{Hiroki Kawabe}
\author{Kaito Ohtani}
\author{Kentaro Yaji\corref{cor1}}
\cortext[cor1]{Corresponding author.}
\ead{yaji@mech.eng.osaka-u.ac.jp}


\address{Department of Mechanical Engineering, Graduate School of Engineering, The University of Osaka, 2-1, Yamadaoka, Suita, Osaka, 565-0871, Japan}

\author{Ryota Fukunishi}
\author{Akira Ogawara}
\address{Engineering and Development Division, NTT DATA XAM Technologies Corporation, 1-1-16, Osaka, Osaka, 555-0034, Japan}

\begin{abstract} 
This paper presents a topology optimization method for designing two-fluid heat exchangers under turbulent conditions using a Darcy flow-based low-fidelity (LF) model.
The LF model is calibrated against a high-fidelity (HF) model based on the Reynolds-averaged Navier-Stokes (RANS) equations to increase the accuracy of predictions for fluid flow and heat transfer characteristics.
Since the discrepancies between the LF and HF models can be significant, particularly for pressure drops, a multifidelity topology optimization framework is adopted to leverage the strengths of both models.
Using the calibrated LF model, we perform topology optimization for various inlet velocities in the boundary conditions and trade-off parameters in the objective function to obtain diverse optimized designs.
The optimized designs are then evaluated using the HF model to assess their performance with higher accuracy.
The results demonstrate that the optimized designs significantly improve overall heat transfer coefficients while maintaining manageable pressure drops, achieving up to a $22$\% higher performance evaluation criterion (PEC) compared to a reference design enhanced by conventional twisted tape insertion.
The improvements are attributed to the optimized configurations that promote enhanced fluid mixing and increased surface area for heat exchange, yet maintain streamlined flow paths to minimize pressure losses.
Overall, the proposed topology optimization method using the Darcy flow-based LF model proves effective in designing high-performance double pipe heat exchangers, showcasing the potential of the multifidelity approach in overcoming the challenges of optimizing heat exchangers under turbulent flow conditions.
\end{abstract}

%


\end{frontmatter}



\begin{table*}[t]
\centering
\small
\setlength{\fboxrule}{0.6pt}  
\setlength{\fboxsep}{6pt}    

\fbox{%
\begin{minipage}{0.98\textwidth}
\textbf{Nomenclature}\par\vspace{8pt}

\renewcommand{\arraystretch}{1.05}

\noindent
\begin{minipage}[t]{0.48\linewidth}
\begin{tabularx}{\linewidth}{@{} l Y @{}}
\multicolumn{2}{@{}l@{}}{\textit{Physics}}\\
$\mb{x}$ & Spatial coordinate \\
$\mb{u}$ & Velocity vector \\
$p$ & Pressure \\
$T$ & Temperature \\
$k$ & Turbulent kinetic energy \\
$\omega$ & Specific dissipation rate \\
$\mu$ & Dynamic viscosity \\
$\mu_\text{t}$ & Turbulent eddy viscosity \\
$\rho$ & Density \\
$c_\text{p}$ & Specific heat at constant pressure \\
$k_{i}$ & Thermal conductivity \\
$C_k$ & Dimensionless thermal conductivity \\
$\kappa$ & Permeability \\
Pe & Péclet number \\
Re & Reynolds number \\
Pr & Prandtl number \\
$\text{Pr}_\text{t}$ & Turbulent Prandtl number \\
Nu & Nusselt number \\
f & Fanning friction factor \\
$h$ & Convective heat transfer coefficient \\
$D_\text{h}$ & Hydraulic diameter \\
$\mb{n}$ & Unit normal vector \\
$\Delta p$ & Pressure drop \\
$L$ & Length of heat exchanger \\
$l$ & Distance from pipe center to wall \\
$t_\text{t}$ & Thickness of twisted tape \\
$w_\text{t}$ & Width of twisted tape \\
$\Delta T_\text{lm}$ & Log mean temperature difference \\
$q$ & Heat flux \\
$U$ & Overall heat transfer coefficient \\
PEC & Performance evaluation criterion \\
\end{tabularx}
\end{minipage}
\hfill
\begin{minipage}[t]{0.48\linewidth}
\begin{tabularx}{\linewidth}{@{} l Y @{}}
\multicolumn{2}{@{}l@{}}{\textit{Optimization}}\\
$\gamma$ & Design variable \\
$\tilde{\gamma}$ & Filtered design variable \\
$\hat{\tilde{\gamma}} $ & Projected design variable \\
$\mb{\xi}$ & Spatial coordinate for filtering \\
$w$ & Weight function of filtering \\
$r$ & Filter radius \\
$\beta$ & Sharpness parameter of projection \\
$\eta$ & Threshold parameter of projection \\
$J$ & Objective function \\
$G_i$ & Constraint functions \\
$R_j$ & Residual functions \\
$\boldsymbol{s}$ & Set of seeding parameters \\
$w_p$ & Weight function of pressure drop \\
$q_i$ & Convexity of permeability interpolation \\
$\sigma$ & Sharpness of thermal conductivity interpolation \\[6pt]

\multicolumn{2}{@{}l@{}}{\textit{Subscripts}}\\
$\text{to}$ & Design optimized by topology optimization \\
in & Inlet \\
out & Outlet \\
total & Total of both fluids \\
$0$ & Smooth pipe reference design \\[6pt]

\multicolumn{2}{@{}l@{}}{\textit{Superscripts}}\\
$\prime$ & Dimensional value \\
$1$ & Fluid 1 \\
$2$ & Fluid 2 \\
$s$ & Solid wall \\
$l$ & Lower bound \\
$u$ & Upper bound \\
$(k)$ & Index of optimization settings \\
$\text{max}$ & Maximum value \\
\end{tabularx}
\end{minipage}

\end{minipage}%
}
\end{table*}

\section{Introduction}
\label{sec:intr}

Heat exchangers (HXs) are vital devices in various engineering applications that facilitate the transfer of thermal energy between two or more fluids at different temperatures through solid materials.
HXs are categorized into many types based on their configurations and working principles, such as plate fin~\cite{zhang2019review}, shell and tube~\cite{marzouk2023comprehensive}, and double pipe HXs (DPHXs)~\cite{tavousi2023heat}.

As the demand for more effcient thermal management systems escalates across industries, each type of HX has been researched extensively to enhance its performance and efficiency since the early 20th century.
For instance, DPHXs are one of the most common HX types, which consist of two concentric pipes allowing two fluids to flow in either parallel or counter flow arrangements, separated by a solid wall.
The most important perspective in enhancing the heat transfer performance of DPHXs is to increase turbulence in the flowing fluids and surface area for heat exchange.
Currently, various techniques have been proposed to achieve this, such as inserting twisted tapes~\cite{dandoutiya2022wcut, hazbehian2016experimental, barzegar2019numerical}, adding fins~\cite{elmaakoul2017numerical}, and employing corrugated tubes~\cite{moyarico2022experimental}.
These techniques, however, are limited in their ability to explore innovative configurations due to the constraints imposed by conventional manufacturing methods, such as bending and twisting tubes.

To overcome these limitations seen in traditional HX designs, additive manufacturing (AM) has been increasingly adopted in recent years~\cite{careri2023additive}.
AM enables the fabrication of complex geometries that are otherwise unachievable with conventional manufacturing techniques, thus opening new avenues for HX designs, which potentially lead to significant improvements in performance with higher degrees of freedom.

For fully leveraging the capabilities of AM, topology optimization (TO), a powerful computational tool for generating innovative HX designs, has gained significant attention.
The basic concept of TO involves converting the original structural optimization problem into a material distribution problem within a fixed design domain, where the presence or absence of material is represented by a characteristic function~\cite{bendsoe1988generating}.
TO was originally developed for solid mechanics problems, but due to its versatility, it has since been extended to various physical problems, including fluid dynamics~\cite{bendsoe2003topology}.

Borrvall and Petersson pioneered the methodology of topology optimization for fluid problems in 2003, applying it to Stokes flow problems~\cite{borrvall2003topology}.
Subsequently, Gersborg-Hansen et al. extended this approach to Navier-Stokes laminar flow problems~\cite{gersborg2005topology}, and further applications to fluidic devices were explored by Lin et al.~\cite{lin2015topology}.
In terms of thermal-fluid problems, TO has been applied to a variety of scenarios, such as forced convection problems~\cite{matsumori2013topology}, natural convection problems~\cite{alexandersen2014topology}, turbulent one-fluid heat transfer problems~\cite{dilgen2018density}, and experimental investigations~\cite{li2019optimal}.
The tremendous progress in TO for fluid problems is comprehensively summarized in the review paper by Alexandersen and Andreasen~\cite{alexandersen2020review}.

Despite the significant advancements, most studies have focused on scenarios involving a single fluid phase interacting with solid structures, unlike HXs that typically involve two distinct fluid phases separated by a solid wall.
Addressing this gap, several studies have explored TO methods specifically tailored for two-fluid heat exchangers.
Papazoglou~\cite{papazoglou2015topology} pioneered the exploration of TO specifically for the design of two-fluid heat exchangers, applying it to density-based TO, one of the most popular TO methods.
The findings underscored the effectiveness of a multi-material model, solving two seperate flow fileds with each of them represented by combination of two distinct design variable fields.
It is, however, noteworthy that the practical applicability of the results is constrained by the assumption of a zero thickness wall, which presents an unrealistic scenario.
Tawk et al.~\cite{tawk2019topology} proposed a TO method for forced convection problems involving domains with two fluids and one solid. 
As a key idea, a penalty function is incorporated to mitigate fluid mixing by introducing a non-zero thickness wall. 
Despite the merits of the approach, the authors acknowledged potential instability in the solution search under certain conditions due to the penalty scheme.
Kobayashi et al.~\cite{kobayashi2020topology} used a single design variable field to represent two fluids and a solid wall, thickness of which is weakly controlled by predefined parameters of filtering and projection techniques.
Similarly, H{\o}gh{\o}j et al.~\cite{hoghoj2020topology} employed a single design variable field to represent three phases, but introduced erosion-dilation techniques \cite{clausen2017topology} to control the wall thickness more explicitly.
Feppon et al.~\cite{feppon2021body} proposed a level set-based TO method, which is another popular TO method, where each phase is represented by a dedicated body-fitted mesh, and the wall thickness is directly controlled by the level set function.
Galanos et al.~\cite{galanos2022method} developed a turbulent two-fluid TO method, followed by a shape optimization using B-spline morphing.
To more accurately capture the fluid-solid interface, Galanos et al.~\cite{galanos2025continuous} developed a cut-cell-based TO method.
Sun et al.~\cite{sun2025three} integrated manufacturing constraints into the density-based TO method to ensure the minimal wall thickness and the overhang angle suitable for AM processes, followed by their experimental work applied to a high temperature HX~\cite{sun2025topology}.
Pimanov et al.~\cite{pimanov2026sparse} proposed a sparse narrow-band TO method for large-scale thermal-fluid applications to focus the computational effort on the fluid-solid interface, where it is most needed.

Most of these preceding studies focused on laminar flow regimes with low Reynolds numbers, except for the studies by Galanos et al.~\cite{galanos2022method,galanos2025continuous}, which utilized Spalart-Allmaras (SA) turbulence models, a type of Reynolds-Averaged Navier-Stokes (RANS) models to handle turbulent flow conditions.
However, even in these studies, the Reynolds numbers considered were relatively low (up to $2560$, based on the inlet length and velocity), leaving the exploration of turbulent flow regimes largely unaddressed.
Turbulent flow has been widely recognized as a key factor in enhancing heat transfer performance in practical HX applications under high Reynolds numbers.
For instance, when twisted tape is inserted in DPHXs, the Reynolds numbers can reach up to $10^4$ or higher and turbulence is deliberately induced to enhance heat transfer~\cite{noorbakhsh2020numerical}.
Therefore, there is a pressing need to develop TO methods that can effectively handle two-fluid HXs under turbulent conditions with high Reynolds numbers.

Despite the development of various forward simulation methods for turbulent flow in fluid mechanics, incorporating them into TO frameworks remains a challenging task due to the increased complexity and computational cost, not only for HXs but also for fluid problems in general.
TO requires numerous iterations of forward simulations to search for optimal configurations, making it computationally prohibitive to directly apply Direct Numerical Simulation (DNS) or Large Eddy Simulation (LES) models.
Moreover, the previous study on turbulent one-fluid heat transfer problems~\cite{dilgen2018density} reported that even when using RANS models, it is inherently difficult or impossible for the density-based TO method to capture the flow behavior near the wall accurately, due to the lack of clearly defined walls during the optimization process.
Furthermore, for both the density-based and level set-based TO methods, resolving the RANS-based TO problems typically requires extremely fine meshes to capture the turbulent eddies and boundary layers accurately, leading to high computational costs.
These challenges can be even more prominent in two-fluid scenarios, where the interaction between two turbulent fluids adds further complexity to the optimization problem.

To address these challenges, Zhao et al.~\cite{zhao2018poor} proposed a density-based TO method for one-fluid turbulent heat transfer problems using a Darcy flow model as a low-fidelity (LF) model instead of a turbulence model, called the \textit{poor man's approach}.
In this method, the optimization problem is formulated using the Darcy flow model, which is computationally less expensive and more stable than turbulence models due to its linear nature.
After optimizing the material distribution by the Darcy flow model, i.e., out of the optimization loop, more accurate evaluations are performed using a turbulence model to assess the performance of the optimized design.
This simple yet effective and versatile framework has been successfully applied to turbulent forced convection problems~\cite{kambampati2020level} and natural convection problems~\cite{asmussen2019poor,pollini2020poor}.

As a consequence of the significant simplification of the flow model, however, discrepancies between the low-fidelity (LF) Darcy model and the high-fidelity (HF) turbulence model are inevitable.
To mitigate the discrepancies between the two models, Kambampati and Kim revealed that lowering global volume fractions of the fluid regions or local volume fractions of the flow channels, i.e., maximum length scale of the flow channels, can improve the agreement between the two models in one-fluid turbulent heat transfer problems~\cite{kambampati2020level}.
This is because the Darcy flow model can not accurately capture the flow behavior like recirculation and separation, which are less likely to occur in narrower flow channels.
This strategy, however, is not directly applicable to two-fluid problems, as narrowing the flow channel of one fluid inherently widens that of the other fluid.

To indirectly address the discrepancies between the two models in two-fluid problems, the study by Galanos et al.~\cite{galanos2022method} proposed a TO framework that initilizes the design variable field using the Darcy flow model, and then updates it using the SA turbulence model.
This two-step approach helps to correct the discrepancies by updating the initial design obtained from the Darcy flow model with the more accurate turbulence model.
However, this approach still relies on the turbulence model within the optimization loop, which can be computationally expensive and may face convergence issues due to the nonlinearity of the turbulence model, escpecially under high Reynolds number conditions.
In addition, since the SA model is a one-equation turbulence model, it tends to be less accurate in capturing complex turbulent flow behaviors inside flow channels, such as strong flow separation and recirculation, which are often observed in practical HX applications.
In fact, this study only demonstrated a benchmark case with the low Reynolds number, without providing the estimation of performance improvement by comparing the optimized designs with conventional HX designs.

Another solution for the discrepancies between the two models is a multifidelity TO framework, where the original optimization problem is indirectly solved by LF optimization using a simplified model and HF evaluation using more accurate model.
The key idea of this framework is to introduce seeding parameters, such as inlet velocities, and to perform multiple independent optimizations differentiated by the seeding parameters using the LF model.
This allows us to obtain a diverse set of optimized designs, which are then evaluated using the HF model to select the most promising design based on the evaluation results (out of the optimization loop).
Yaji et al.~\cite{yaji2020multifidelity} proposed a multifidelity framework for one-fluid turbulent heat transfer problems, where the LF and HF model are based on the laminar flow model and the RANS model, respectively.
Ohtani et al.~\cite{ohtani2025multifidelity} used the same concept in one-fluid turbulent heat transfer problems, but adopted the Darcy flow model as the LF model instead, and demonstrated the effectiveness of the framework in obtaining high-performance designs with low computational costs.
Yuan et al.~\cite{yuan2026multifidelity} applied the multifidelity TO framework to a boiling heat transfer problem, proving its versatility in tackling optimization problems that are difficult to solve directly due to strong nonlinearities.

However, despite the recent progress in TO for HXs, three limitations remain.
First, most two-fluid heat-exchanger studies have focused on laminar or relatively low-Reynolds-number conditions.
Second, directly incorporating turbulence models into the optimization loop remains computationally expensive and numerically challenging, especially for two-fluid problems.
Third, the practical benefit of turbulent two-fluid TO has not been sufficiently demonstrated through quantitative comparison with conventional enhancement techniques.
To address these limitations, this study makes the following contributions.
First, we develop a multifidelity topology optimization framework for two-fluid heat exchangers under turbulent conditions, where a Darcy-flow-based LF model is used for optimization and a RANS-based HF model is used only for evaluation.
Second, we introduce a calibration strategy for the LF Darcy model to improve its consistency with the HF response in two-fluid turbulent heat-transfer problems, with the HF response validated against empirical correlations.
Third, we apply the proposed framework to DPHXs over Reynolds numbers up to 13,000 and demonstrate, through HF analysis, that the optimized designs can outperform a conventional twisted-tape reference in terms of thermohydraulic performance, achieving up to 22\% higher PEC.

The rest of the paper is organized as follows. In Section 2, we present the overall framework of the proposed method, including the Darcy flow model-based TO and the subsequent HF evaluation using the RANS model.
In Section 3, we detail the formulation of the Darcy flow model and the TO problem.
In Section 4, we present numerical examples of DPHX designs using the proposed method and discuss the results.
Finally, in Section 5, we summarize the findings and discuss future research directions.

\section{Framework}
\label{sec:fram}
In this section, we introduce the overall framework of the proposed method for designing two-fluid heat exchangers under turbulent conditions.

\subsection{Topology optimization}
\label{subsec:top}

\begin{figure*}[h]
\centering
\includegraphics[width=\textwidth]{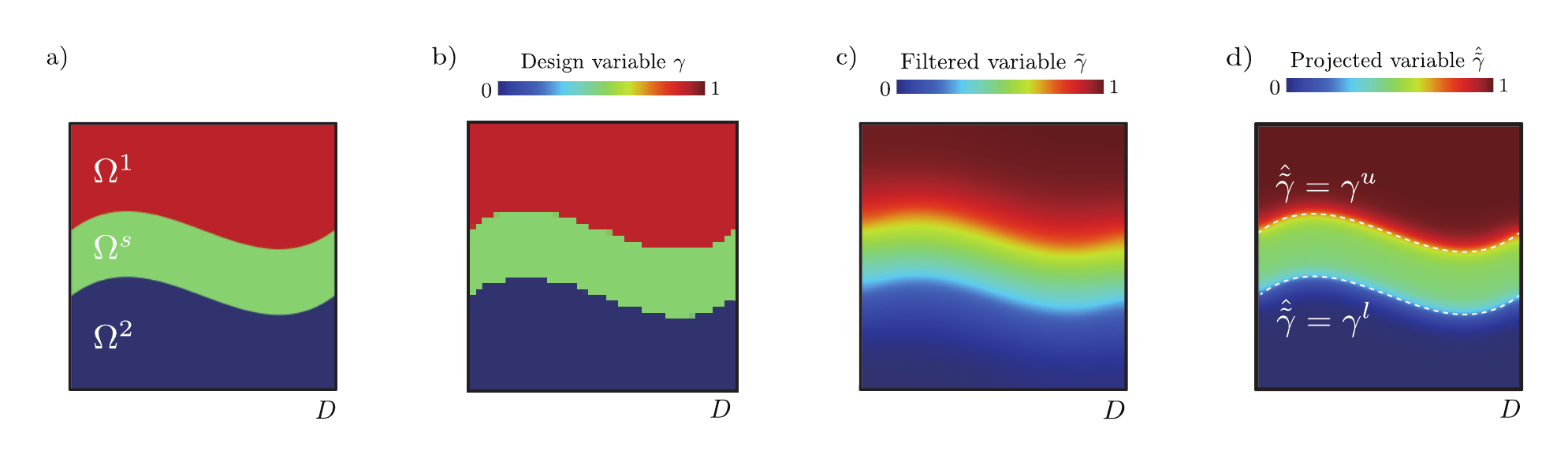}
\caption{Sketch of the subdomains in the design domain $D$: a) original subdomains $\Omega^1, \Omega^2, \Omega^s$; b) design variable field $\gamma$; c) filtered design variable field $\tilde{\gamma}$; d) projected design variable field $\hat{\tilde{\gamma}}$.}
\label{fig:domain_concept}
\end{figure*}

The ultimate goal of the optimization problem is to optimize the configuration of the two fluids and the separating wall, focusing on the trade-off between heat transfer and pressure loss.
This problem targets a fixed design domain $D\subset \mathbb{R}^d$ with $d\in\{2,3\}$ that encompasses the entire region of interest, including the fluid domains $\Omega^1, \Omega^2$ and the solid wall domain $\Omega^s$, as illustrated in Fig.~\ref{fig:domain_concept}a.
Since the original problem is inherently discrete and impractical to solve directly, we employ density approach~\cite{bendsoe2003topology} to relax the optimization problem into a continuous one.
Specifically, we use a single design variable field $\gamma: D\rightarrow[0,1]$ to represent the three subdomains, and discretize the design variable field into a finite number $N$ of design variables, as shown in Fig.~\ref{fig:domain_concept}b.
This relaxation allows us to formulate the optimization problem in a continuous manner, enabling the use of gradient-based optimization techniques for efficient solution search.

To ensure the spatial smoothness of the design variable $\gamma$, the density filter~\cite{bourdin2001filters} is applied to obtain the filtered design variable $\tilde{\gamma}$:

\begin{align}
  w(\mathbf{x},\boldsymbol{\xi}) =
  \begin{cases}
    \dfrac{r - \lVert \mathbf{x}-\boldsymbol{\xi} \rVert}{r}, & \text{if } \lVert \mathbf{x}-\boldsymbol{\xi} \rVert < r, \\[6pt]
    \quad \quad \ 0, & \text{otherwise}.
  \end{cases}
\label{eq:weight}
\end{align}

\begin{align}
  \tilde{\gamma}(\mathbf{x})
  =
  \frac{\displaystyle
    \int_{\Omega} w(\mathbf{x},\boldsymbol{\xi})\,
                   \gamma(\boldsymbol{\xi})\,\mathrm{d}\boldsymbol{\xi}}
       {\displaystyle
    \int_{\Omega} w(\mathbf{x},\boldsymbol{\xi})\,
                   \mathrm{d}\boldsymbol{\xi}},
  \label{eq:density_filter}
\end{align}
\noindent where $r$ is the filter radius parameter; $w(\mb{x}, \boldsymbol{\xi})$ is the weight function defined as a hat-type kernel.
The filtered design variable $\tilde{\gamma}$ is shown in Fig.~\ref{fig:domain_concept}c.

After filtering, the filtered design variable $\tilde{\gamma}$ is projected to further reduce intermediate grayscale values between 0 and 1, using the Heaviside projection~\cite{kawamoto2011heaviside}:
\begin{align}
  \hat{\tilde{\gamma}}=\frac{\tanh(\beta\eta)+\tanh(\beta(\tilde{\gamma}-\eta))}{\tanh(\beta\eta)+\tanh(\beta(1-\eta))},
  \label{eq:projection}
\end{align}
\noindent where $\beta$ and $\eta$ serve as the sharpness and the threshold parameters to control the shape of the Heaviside function, respectively.
The projected design variable $\hat{\tilde{\gamma}}$ is illustrated in Fig.~\ref{fig:domain_concept}d.

After obtaining the projected design variable $\hat{\tilde{\gamma}}$, the three subdomains are extracted as follows:
\begin{align}
  \begin{aligned}
  \Omega^1 &= \{\mb{x}\in D \mid \hat{\tilde{\gamma}}(\mb{x}) \leq \gamma^{l}\},\\
  \Omega^s &= \{\mb{x}\in D \mid \gamma^{l} < \hat{\tilde{\gamma}}(\mb{x}) < \gamma^{u}\},\\
  \Omega^2 &= \{\mb{x}\in D \mid \hat{\tilde{\gamma}}(\mb{x}) \geq \gamma^{u}\},
  \end{aligned}
  \label{eq:subdomains}
\end{align}
\noindent where $\gamma^{u}$ and $\gamma^{l}$ are the predefined threshold values to separate the three subdomains.
Those threshold values are chosen such that, in the reference configuration of the projected design variable $\hat{\tilde{\gamma}}$, the volume of the extracted solid wall domain $\Omega^s$ matches that of the desired wall thickness, satisfying $0 < \gamma^{l} < \gamma^{u} < 1$.
This single design variable field approach can ensure both fluid domains are always separated by a solid wall with a non-zero thickness, which is essential for two-fluid heat exchangers.

\subsection{Multifidelity approach}
\label{subsec:highfidelity}
Despite its efficient solution search capability proven by extensive studies, TO often faces challenges associated with complex solution spaces, called multimodality, leading to numerous local optima.
These challenges, in particular, become more prominent with strongly nonlinear physical fields, such as turbulent flows.
This often necessitates extensive parameter studies to obtain promising design solutions, with each study requiring significant computational resources.

To address this issue, we utilize an indirect approach that leverages LF models for the optimization process, followed by HF evaluations of the obtained designs.
Following the methodology proposed by Zhao et al.~\cite{zhao2018poor}, in the optimization process, we substitue the Navier-Stokes equations, known for their numerical instabilities, with the Darcy flow equations, which are computationally less expensive and more stable.

\begin{figure}[t]
  \centering
  \includegraphics[width=0.8\columnwidth]{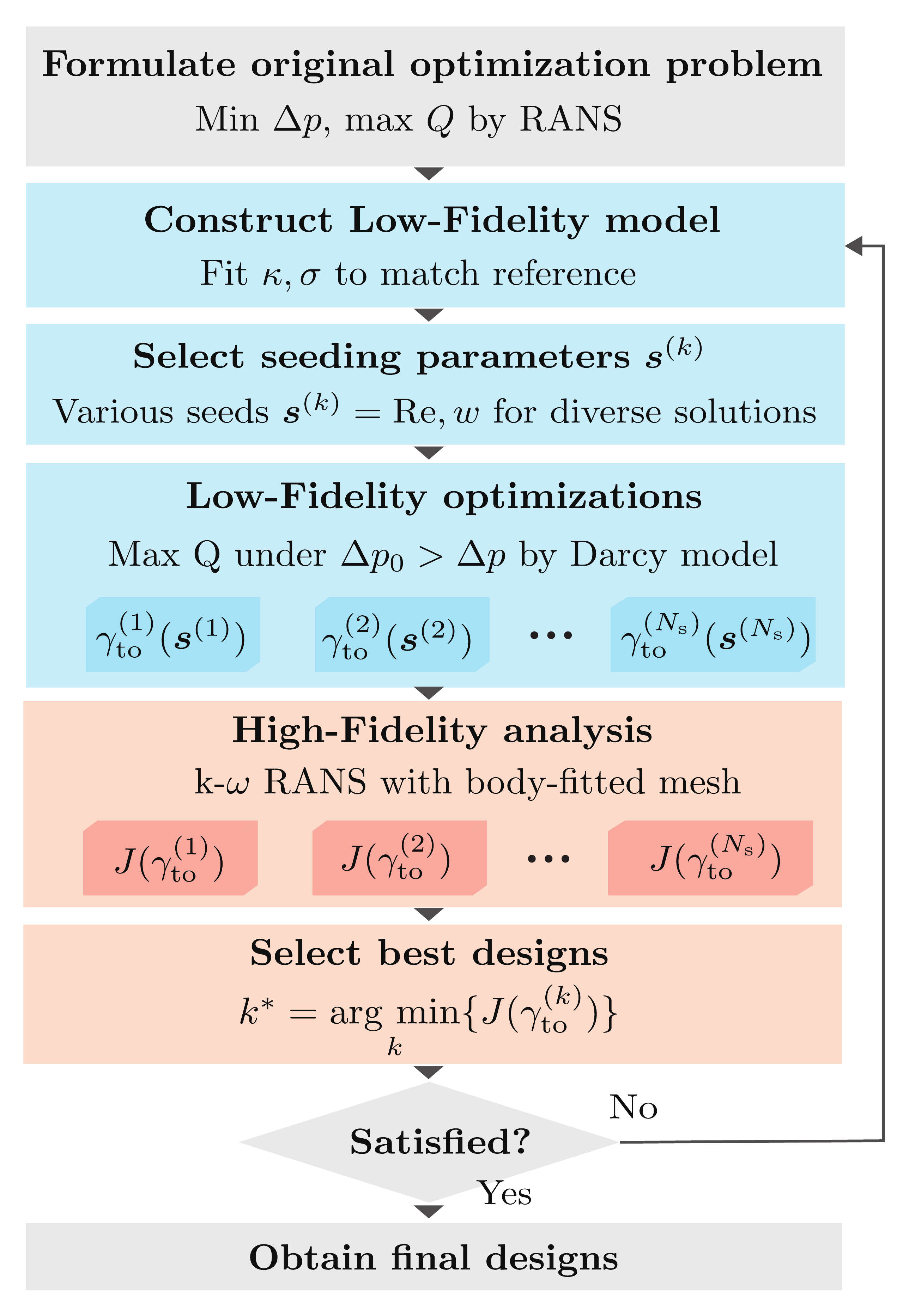}
  \caption{Schematic of the proposed framework.}
  \label{fig:framework}
  \end{figure}

Schematic of the proposed framework is illustrated in Fig~\ref{fig:framework}.
The initial step involves fitting the LF model to closely match the HF model in the reference configuration, in terms of velocity and temperature fields.
The temperature is fitted by adjusting the thermal parameters of the LF model so that the heat transfer  of both models align as closely as possible, whereas the velocity is fitted by tuning the permeability parameters of the LF model to minimize the discrepancy in pressure drops between the two models.
After fitting, the second step is to perform TO using the pre-fitted LF model, for a variety of optimization settings, such as different velocities of inlet flows in the governing equations and different weightings between heat transfer and pressure loss in the objective function, to generate diverse candidate designs.
A general formulation of the LF model is given as follows:
\begin{equation}
    \begin{aligned}
    \makebox[2cm][l]{$\underset{\gamma}{\text{minimize}}$}   & \tilde{J} \big(\gamma^{(k)}, \boldsymbol {s}^{(k)} \big) \\
    \makebox[2cm][l]{subject to} & \tilde{G}_i \big(\gamma^{(k)}, \boldsymbol {s}^{(k)} \big) \leq 0, \quad \quad \ i = 1, 2, ..., N_c, \\
                                & 0 \leq \gamma^{(k)}(\boldsymbol{x}) \leq 1, \quad \quad \ \ \forall \boldsymbol{x} \in D, \\
    \makebox[2cm][l]{where} & \tilde{R}_j \big(\gamma^{(k)}(x),\boldsymbol{s}^{(k)}\big)=0, \quad j=1, 2, ..., N_r \\
    \makebox[2cm][l]{for given}  & \boldsymbol {s}^{(k)} \in S, \quad \quad \quad \quad \quad \ \ \ k=1, 2, ..., N_s,
    \end{aligned}
\label{eq:low-fidelity}
\end{equation}
\noindent where $\tilde{J}$ and $\tilde{G}_i$ are the objective function and the constraint functions defined on the LF model, respectively; $\tilde{R}_j$ are the governing equations of the LF model; $\boldsymbol{s}^{(k)}$ is the set of parameters defining the $k$-th optimization setting; $S$ is the set of all optimization settings; $N_s$ is the total number of optimization settings.

Since the flow model is largely simplified to the Darcy flow model, the optimized design does not necessarily perform well on the HF model.
Therefore, the third step is to extract clear boundaries between the three regions from the LF optimization results, as explained in the last subsection, and evaluate the heat transfer performance and pressure loss of the optimized designs on the HF model using a turbulence model with wall functions.
The evaluated designs from all optimization settings are compared based on the HF objective function, and the best design is selected as follows:
\begin{equation}
k^* = \underset{k}{\text{arg min}} \{ J(\gamma_{\text{to}}^{(k)}) \ | \ G_i (\gamma_{\text{to}}^{(k)}) \leq 0, \ R_j (\gamma_{\text{to}}^{(k)}) = 0\}.
\label{eq:high-fidelity}
\end{equation}
\noindent where $J$ is the objective function accurately evaluated on the HF model; $G_i$ are the constraint functions of the HF model; $R_j$ are the residuals of the governing equations of the HF model; $\gamma_{\text{to}}^{(k)}$ is the optimized design obtained from the LF optimization under the $k$-th setting; $k^*$ is the index of the best design among all candidates evaluated on the HF model.

When the evaluated designs do not meet the desired performance, in other words, when no design outperform the baseline design in terms of the HF objective function $J$, modifications to the LF optimization model may be necessary, such as refining the fitting process or exploring additional optimization settings $\boldsymbol{s}^{(k)}$.
By iteratively refining the LF model and exploring various optimization settings, we can enhance the likelihood of obtaining designs that perform well on the HF model, making it possible to tackle optimization problems that are otherwise challenging or infeasible to solve directly.

\section{Formulation}
\label{sec:for}
This section provides the detailed formulation of our proposed method for TO problems of two-fluid heat exchangers under turbulent conditions.
We first elaborate a Darcy flow model with convection heat transfer to be used as the LF model in the optimization process.
Next, we formulate the LF optimization problem using the proposed Darcy flow-based model.
We then show the detailed procedures for the HF evaluation concerning a turbulent heat transfer.

\subsection{Darcy flow model for two-fluid heat exchangers}
\label{subsec:dar}

As explained in the section~\ref{subsec:top}, the solid wall domain $\Omega^s$ is explicitly extracted within $\gamma^{l} < \hat{\tilde{\gamma}} < \gamma^{u}$ after the optimization, to be used as the HF evaluation model.
In contrast, during the optimization process, the continous representation of the three subdomains is maintained using two distinct types of interpolation functions in the governing equations of the flow and temperature fields, thereby ensuring numerical stability.
All parameters and variables in the LF Darcy flow model are non-dimensionalized for numerical stability, as detailed in~\ref{ap:dimensionless}.

For the flow field, the Darcy flow model is formulated with continuity equations and the Darcy equations for each fluid, as follows:
\begin{align}
  &\nabla\cdot\mb{u}_i=0,
  \label{eq:continuous}
  \\
  &\mb{u}_i=- \frac{\kappa_i}{\mu_i} \nabla p_i,
  \label{eq:momentum}
\end{align}
\noindent where $i\in\{1,2\}$ is the index for fluid 1 and fluid 2; $\mb{u}_i$ is velocity; $p_i$ is pressure; $\mu_i$ is viscosity; $\kappa_i$ is permeability, which depends on the projected design variable field $\hat{\tilde{\gamma}}$ to represent the fluid and solid states.
The pearmiabilities for the two fluids, $\kappa_1$ and $\kappa_2$, are defined using the RAMP interpolation functions~\cite{stolpe2001alternative,alexandersen2023detailed} as follows:
\begin{align}
    &\displaystyle \kappa_1 = \kappa^{\max}_1 \frac{\hat{\tilde{\gamma}}}{1+q_1(1-\hat{\tilde{\gamma}})},
    \label{eq:kappa1}
    \\
    &\displaystyle \kappa_2 = \kappa^{\max}_2 \frac{1-\hat{\tilde{\gamma}}}{1+q_2\hat{\tilde{\gamma}}},
    \label{eq:kappa2}
\end{align}
\noindent where $\kappa^{\max}_i$ is the maximum value of the permeability for each fluid; $q_i$ is the parameter to control the convexity of $\kappa_i$.

\begin{figure}[t]
\centering
\includegraphics[width=\columnwidth]{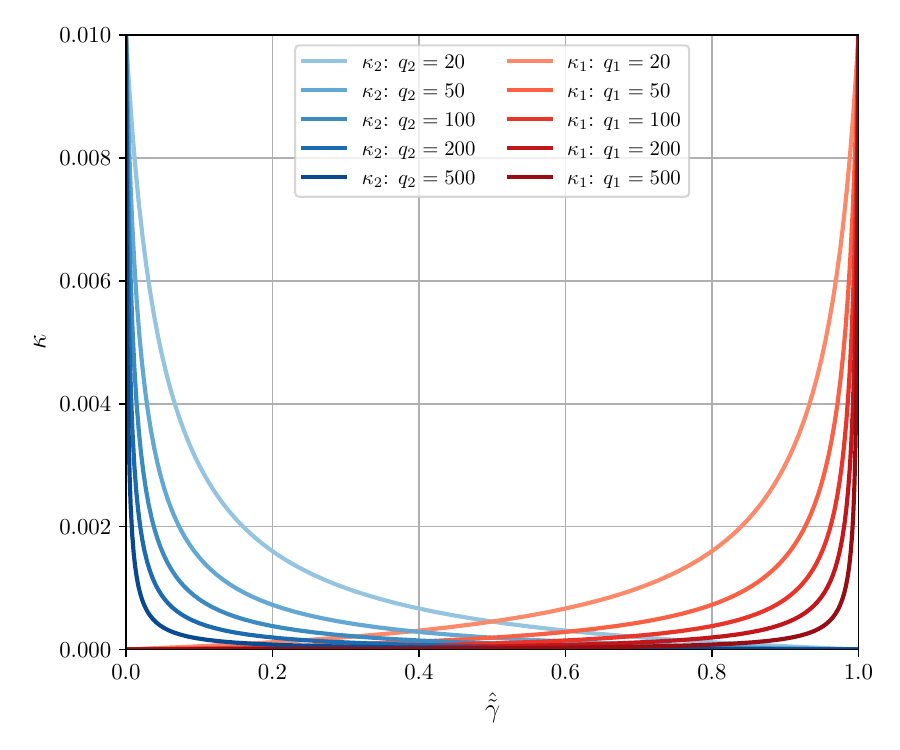}
\caption{Interpolation function for permeability.}
\label{fig:permeability}
\end{figure}

Fig.~\ref{fig:permeability} draws the interpolation functions at $\kappa^{\max}_1=\kappa^{\max}_2=0.01$ and $q_1=q_2=20, 50, 100, 200, 500$.
This figure illustrates a symmetrical relationship between the two permeabilities with respect to $\hat{\tilde{\gamma}}$.
This symmetrical property ensures that when $\hat{\tilde{\gamma}}=0$, $\kappa_1$ attains its maximum value, $\kappa^{\max}_1$, while $\kappa_2$ becomes zero, allowing only fluid 1 to flow while fluid 2 remains immobile.
By solving Eqs.~(\ref{eq:continuous}) and (\ref{eq:momentum}) with these permeability definitions individually for each fluid, we can differentiate the fluid region and the ``solid'' region that includes the solid wall and the opposite fluid region.
As shown in Fig.~\ref{fig:permeability}, increasing the values of $q_1$ and $q_2$ sharpens the transition of the permeabilities between the fluid and solid states.
To effectively represent the wall between the two fluids, the parameters $q_1$ and $q_2$ should be chosen such that both permeabilities $\kappa_1$ and $\kappa_2$ are sufficiently small in the intermediate range of $\hat{\tilde{\gamma}}$, thereby significantly restricting the flow of both fluids.

\begin{figure}[t]
\centering
\includegraphics[width=\columnwidth]{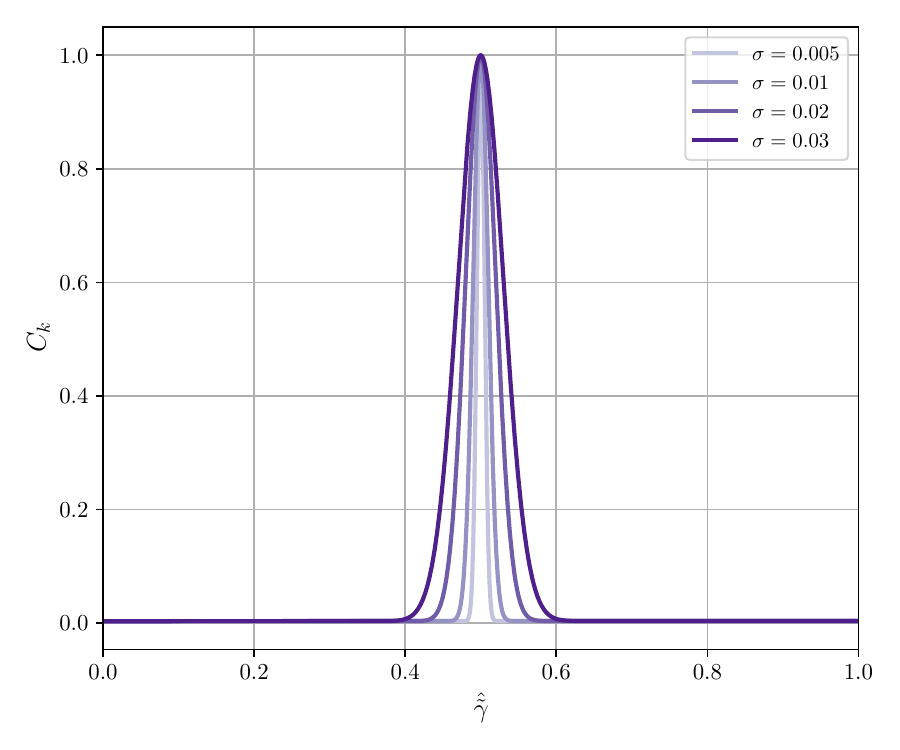}
\caption{Interpolation function for thermal conductivity.}
\label{fig:thermal_conductivity}
\end{figure}

For the forced convection heat transfer problem, a steady-state energy equation is considered, as follows:
\begin{equation}
  (\text{Pe}_1 \mb{u}_1+ \text{Pe}_2 \mb{u}_2)\cdot\nabla T - C_k \nabla^2 T = 0,
  \label{eq:energy}
\end{equation}
\noindent where $\text{Pe}_i$ is the Péclet number; $T$ is the temperature; $C_k$ is the dimensionless thermal conductivity.
The derivation of this equation is provided in~\ref{ap:energy_equation}.
$C_k$ is defined using an interpolation function to represent the three subdomains, using a Gaussian function as follows:
\begin{equation}
  C_k= \frac{1}{k_\text{s}}
  \begin{cases}  
  \left(k_\text{s} - k_2\right) \exp \left(-\frac{(\hat{\tilde{\gamma}}-0.5)^2}{2\sigma^2} \right) + k_2, & (0 \leq\hat{\tilde{\gamma}} \leq 0.5), \\[6pt]
  \left(k_\text{s} - k_1\right) \exp \left(-\frac{(\hat{\tilde{\gamma}}-0.5)^2}{2\sigma^2} \right) + k_1, & (0.5 < \hat{\tilde{\gamma}} \leq 1), \\[6pt]
  \end{cases}
\label{eq:ck}
\end{equation}
\noindent where $k_i$ is the thermal conductivity for each fluid; $k_\text{s}$ is the thermal conductivity for the solid wall; $\sigma$ is the control parameter to determine the sharpness of the Gaussian function.
Fig.~\ref{fig:thermal_conductivity} illustrates the interpolation function of $C_k$ at $k_1=0.654$, $k_2=0.602$, $k_\text{s}=202$ [W/(m$\cdot$K)], and $\sigma=0.005,0.01,0.02,0.03$.
This function smoothly interpolates the thermal conductivity between the three subdomains, ensuring that in the fluid 1 region ($\hat{\tilde{\gamma}}=1$), $C_k k_\text{s}$ equals $k_1$; similarly in the fluid 2 region ($\hat{\tilde{\gamma}}=0$), $C_k k_\text{s}$ equals $k_2$; and in the solid wall region ($\hat{\tilde{\gamma}}=0.5$), $C_k k_\text{s}$ approaches $k_\text{s}$.
As shown in Fig.~\ref{fig:thermal_conductivity}, decreasing the value of $\sigma$ sharpens the transition of $C_k$ between the three subdomains.

The boundary conditions for Eqs.~(\ref{eq:continuous}), (\ref{eq:momentum}), and (\ref{eq:energy}) are defined as follows:
\begin{align}
  &\mb{u}_i=u_{\mathrm{in}, i}, \quad T=T_{\mathrm{in},i} \quad \mathrm{on} \ \Gamma_i^{\mathrm{in}},\label{eq:bc_1}\\
  &p_i=p_{\mathrm{out}, i}, \quad \nabla T \cdot \mathbf{n} = 0 \quad \mathrm{on} \ \Gamma_i^{\mathrm{out}},\label{eq:bc_2}
\end{align}
\noindent where the inlet and outlet for each fluid are denoted as $\Gamma_i^{\mathrm{in}}$ and $\Gamma_i^{\mathrm{out}}$, respectively; $u_{\mathrm{in}, i}$ and $T_{\mathrm{in},i}$ are the prescribed inlet velocity and temperature for each fluid; $p_{\mathrm{out}, i}$ is the prescribed outlet pressure for each fluid; $\mathbf{n}$ is the unit normal vector on the boundary.

\subsection{Optimization problem}
\label{subsec:opt}

Based on the Darcy flow model described in the previous subsection, we now formulate the TO problem for two-fluid heat exchangers.
The objective of this optimization problem is to maximize the total thermal energy transfer between the two fluids, and minimize the pressure drops of both fluids.
The heat flux for each fluid $J_i^h$ and the pressure drop $\Delta p_i$ are defined as follows:
\begin{align}
  \begin{aligned}
    J_i^h & = \frac{\text{Pe}_i}{\int_{\Gamma_i^{\mathrm{out}}} d\Gamma} \int_{\Gamma_i^{\mathrm{out}}} (\mathbf{u}_i \cdot \mathbf{n})(T-T_{\mathrm{in},i}) \ \text{d}\Gamma, \\
    \Delta p_i & = \frac{1}{\int_{\Gamma_i^{\mathrm{in}}} d\Gamma} \int_{\Gamma_i^{\mathrm{in}}} (p_i - p_{\mathrm{out},i}) \ \text{d}\Gamma,
    \label{eq:J}
    \end{aligned}
\end{align}
\noindent where $\mathbf{n}$ is the unit normal vector on the boundary pointing outward from the domain.
The derivation of the heat flux $J_i^h$ is provided in~\ref{ap:heat_flux}.
The optimization problem is then formulated as follows:
\begin{align}
  \begin{aligned}
    & \underset{\gamma({\mathbf{x}})}{\mathrm{maximize}}\quad {J} = J_1^h + J_2^h - w_p \left( \Delta p_1  + \Delta p_2 \right), \\
    & \mathrm{subject\ to}\quad 0 \leqslant \gamma({\mathbf{x}}) \leqslant 1\quad \mathrm{for} \ \forall {\mathbf{x}} \in D,
    \label{eq:optimization_problem_total}
    \end{aligned}
\end{align}
\noindent where $w_p$ is a trade-off parameter to balance the heat transfer and pressure drop.
The first two terms in the objective function represent the total heat transfer between the two fluids, which we aim to maximize.
The last two terms represent the total pressure drop of both fluids, which we aim to minimize.
Here, the pressure drops are included in the objective function rather than in constraints, because we aim not to meet a specific pressure drop target, but to explore various designs with different trade-offs between heat transfer and pressure drop by changing the value of $w_p$, thereby generating a diverse set of candidate designs.

When the trade-off parameter $w_p$ is not sufficiently large, the greyscale region often appears in the optimized design, where two fluids can flow with moderate permeabilities in the same region, leading to impractical designs.
In this case, the heat transfer $J^h$ increases due to an undesired increase in the convective heat transfer in the greyscale region, whereas the pressure drop $\Delta p_i$ increases due to the relatively low permeabilities for both fluids.
Therefore, by gradually increasing the trade-off parameter $w_p$ while solving the optimization problem in Eq.~(\ref{eq:optimization_problem_total}), the optimization process is encouraged to reduce the greyscale region to lower the pressure drop terms in the objective function.
Thus, we can obtain a clear separation between the two fluid regions and the solid wall region in the optimized design.

\subsection{High-fidelity model for turbulent heat transfer}
\label{subsec:hig}

To evaluate the performance of the optimized designs obtained from the LF optimization process, we now formulate the HF model for turbulent heat transfer.
As explained in the section~\ref{sec:fram}, we define the subdomains based on the projected design variable field $\hat{\tilde{\gamma}}$ obtained from the LF optimization, and extract each region, i.e, $\Omega^s, \Omega^1, \Omega^2$, using Eq.~(\ref{eq:subdomains}) to construct a body-fitted domain for the turbulent heat transfer analysis.

To simulate the turbulent flow field, we employ $k-\omega$ model with wall functions, one of the most widely used RANS models~\cite{davidson2004turbulence}, as follows:
\begin{equation}
  \begin{aligned}
  \nabla\cdot\mb{u} &=0, \\
  \rho(\mb{u}\cdot\nabla)\mb{u} &= -\nabla p + \nabla\cdot(\mu+\mu_\text{t})(\nabla \mb{u} + (\nabla \mb{u})^\mathsf{T}),
  \end{aligned}
  \label{eq:RANS}
\end{equation}
\noindent where $\mu_\text{t}$ is the turbulent eddy viscosity:
\begin{align}
\mu_\text{t}=\rho\frac{k}{\omega},
\label{eq:mt}
\end{align}
\noindent where $k$ is the turbulent kinetic energy; and $\omega$ is the specific dissipation rate.
The governing equations for $k$ and $\omega$ are given as follows:
\begin{align}
&\rho(\mb{u}\cdot\nabla)k=\nabla\cdot\left(\left(\mu+\mu_\text{t}\sigma_k^*\right)\nabla k\right)+P_k -\beta_0^*\rho k \omega,
\label{eq:tm_k_w}
\\
&\rho(\mb{u}\cdot\nabla)\omega=\nabla\cdot\left(\left(\mu+\mu_\text{t}\sigma_{\omega}\right)\nabla \omega\right)+ \alpha \frac{\omega}{k} P_k -\rho \beta_0 \omega^2,
\label{eq:tm_w}
\end{align}
\noindent where $P_k$ is the production term of the turbulent kinetic energy defined as $P_k=\mu_\text{t}\left[\nabla\mb{u}:(\nabla\mb{u}+(\nabla\mb{u})^\mathsf{T})\right]$; the empirical constants are set as $\beta_0^*=0.09$, $\beta_0=0.072$, $\alpha=0.52$, $\sigma_k^*=0.5$, and $\sigma_{\omega}=0.5$.

For the temperature field, the energy equation considering turbulent heat transfer is given as follows:
\begin{align}
  \mb{u}\cdot\nabla T=\nabla\cdot\left(\left(\frac{\nu}{\text{Pr}}+\frac{\nu_\text{t}}{\text{Pr}_\text{t}}\right)\nabla T\right),
  \label{eq:tm_ch}
\end{align}
where $\text{Pr}_\text{t}$ is the turbulent Prandtl number; $\nu_\text{t}=\mu_\text{t}/\rho$ is the turbulent kinematic viscosity; $\nu=\mu/\rho$ is the kinematic viscosity; and $\text{Pr}$ is the molecular Prandtl number.

\section{Numerical examples}
\label{sec:num}

\begin{figure}[t]
\centering
\includegraphics[width=\columnwidth]{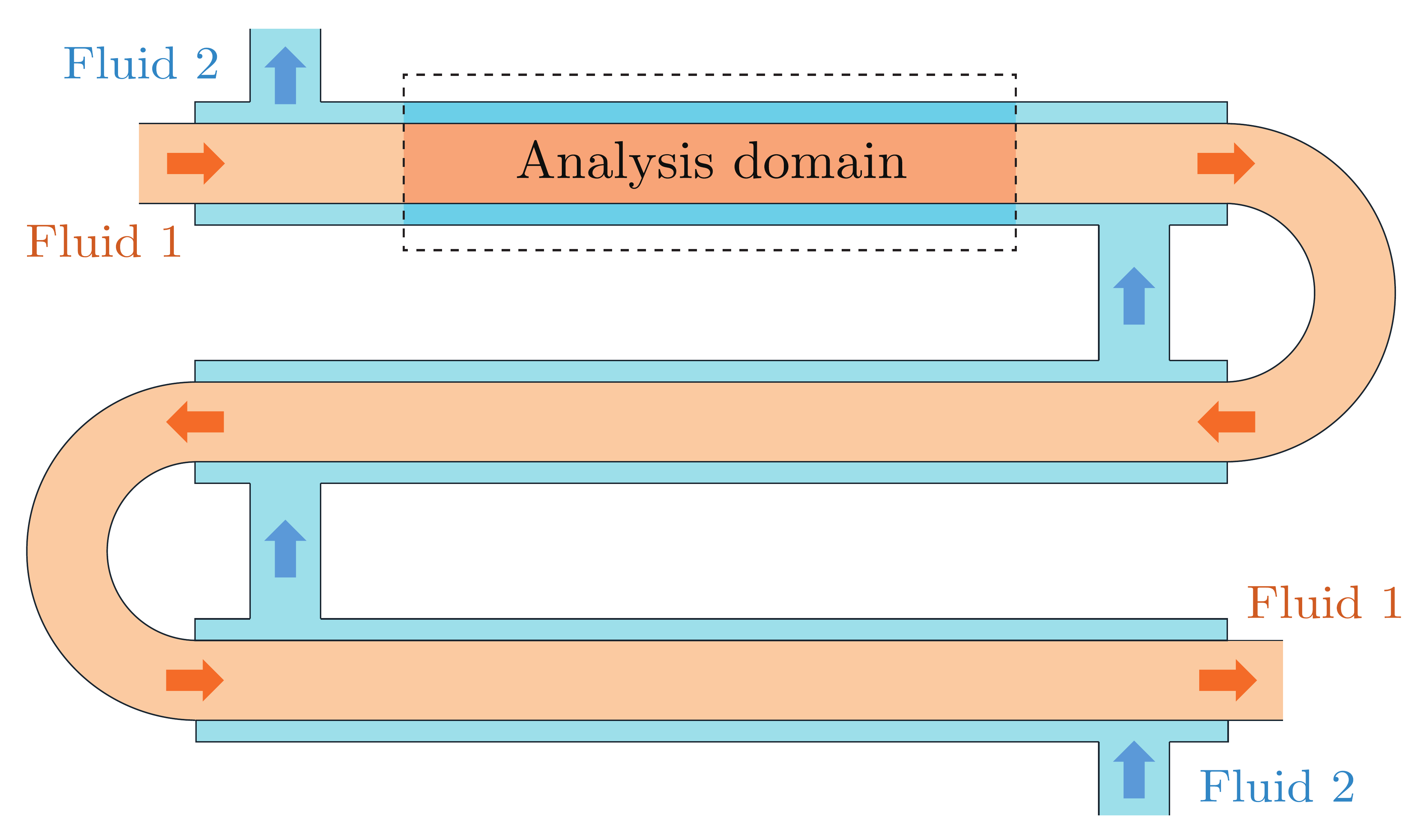}
\caption{Illustration of the DPHX targeted in this study.}
\label{fig:hx}
\end{figure}

This section presents numerical examples to demonstrate the effectiveness of the proposed framework.
We apply the proposed framework to a DPHX, especially targeting its major components, i.e., the core pipes and the solid wall between the two fluids except for the flange sections, as illustrated in Fig.~\ref{fig:hx}.
Note that we focus on demonstrating that the Darcy flow-based TO can effectively generate promising designs for the DPHX, rather than achieving the best practical, manufacturing-ready design with manufacturing constraints of AM.
References on incorporating manufacturing constraints in TO have been extensively published, e.g., self-supported structure design method~\cite{langelaar2017additive}, and these methods can be combined with the proposed framework if necessary.

\subsection{Problem setup}
\label{subsec:prob}

\begin{figure}[t]
\centering
\includegraphics[width=\columnwidth]{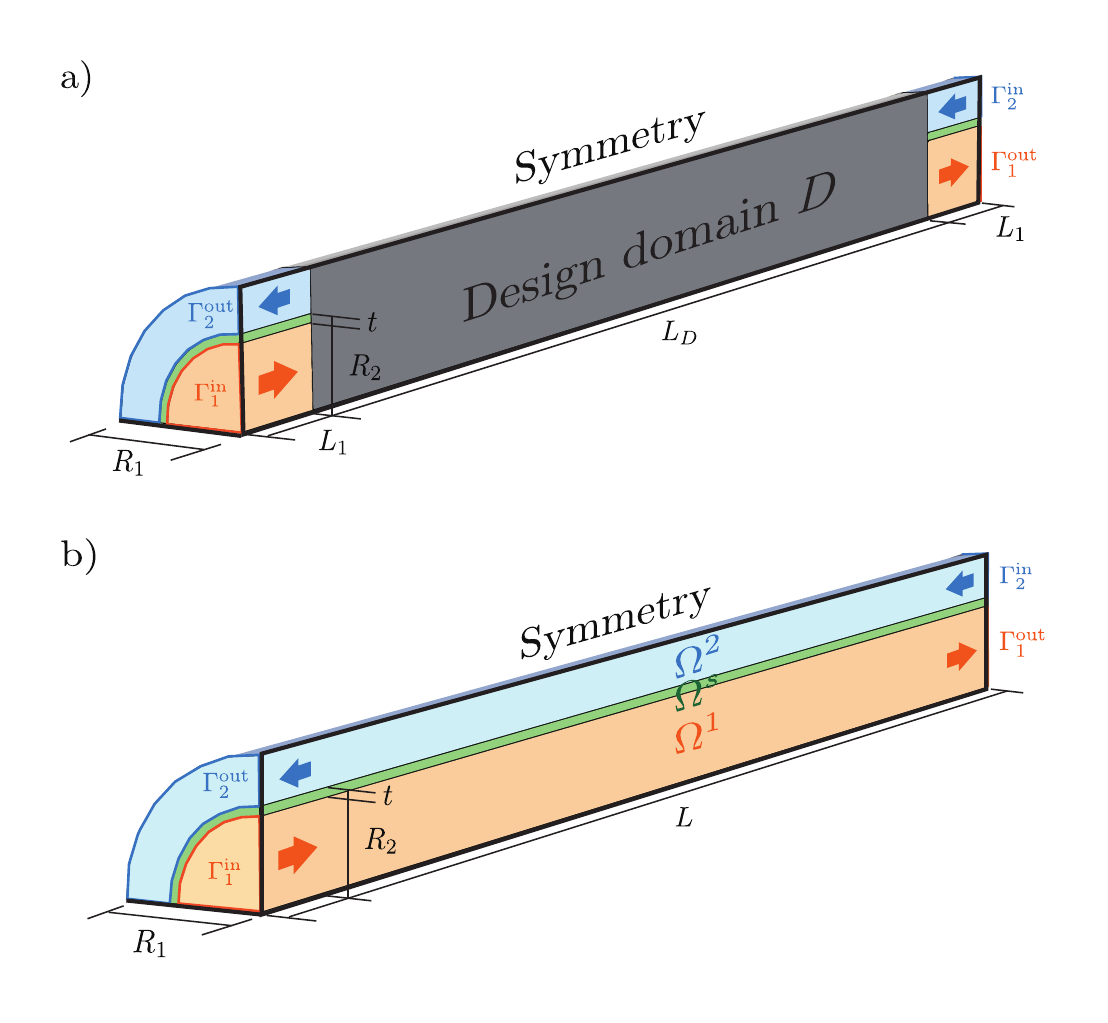}
\caption{Dimensions and boundary conditions: a) LF model; b) HF model.}
\label{fig:setup}
\end{figure}

Fig.~\ref{fig:setup} illustrates the dimensions and boundary conditions of the LF and HF models.
Only quater domain of each region is modeled in both models to reduce the computational costs, by applying symmetry boundary conditions.
In both models, fluid 1 (high-temperature fluid) enters from the left inlet and exits from the right outlet of inner pipe, while fluid 2 (low-temperature fluid) enters from the right inlet and exits from the left outlet of outer pipe.
For LF model (Fig.~\ref{fig:setup}a), the entire domain is composed of the design domain $D$, where the three subdomains are represented using the projected design variable field $\hat{\tilde{\gamma}}$, and non-design domains with fixed solid walls placed at both ends of the design domain.
The boundary conditions for the LF model are defined using Eqs.~(\ref{eq:bc_1}) and (\ref{eq:bc_2}), where the inlet temperatures are set to $T_{\mathrm{in},1}=333.15$ [K] and $T_{\mathrm{in},2}=293.15$ [K], respectively; and the outlet pressures are set to $p_{\mathrm{out},1}=p_{\mathrm{out},2}=0$ [Pa].

Dimensions of heat exchanger model are defined following the previous work~\cite{dandoutiya2022wcut} and are summarized in Table~\ref{tab:dim}.
The physical properties of the hot and cold fluids are defined as those of water at temperatures of $333.15$ K and $293.15$ K, respectively, while the solid wall is assumed to be made of aluminum, as summarized in Table~\ref{tab:prop}.
The inlet velocities for fluid 2 is set to a value of $u_{\mathrm{in},2}=0.144$ m/s.
For fluid 1, five different inlet velocities are considered as $u_{\mathrm{in},1}=$ $0.144$, $0.217$, $0.289$, $0.361$, $0.433$ m/s, corresponding to Reynolds numbers of $\text{Re} = \rho_1 u_{\mathrm{in},1} D_h / \mu_1 = 4{,}300, 6{,}500, 8{,}700, 10{,}900, 13{,}000$, respectively, to cover a wide range of turbulent flow conditions.

\begin{table}[t]
\centering
\caption{Dimensions of the heat exchanger model.}

{\footnotesize
  \renewcommand{\arraystretch}{1}
\begin{tabular}{lc}
\hline
Parameter & Value \\
\hline
Pipe length $L$ & $120$ mm \\
Non-design domain length $L_1$ & $10$ mm \\
Design domain length $L_D$ & $100$ mm \\
Outer pipe diameter $R_1$ & $11.7$ mm \\
Inner pipe diameter $R_2$ & $7.15$ mm \\
Wall thickness $t$ & $0.75$ mm \\
Thickness of twisted tape $t_{\mathrm{t}}$ & $2$ mm \\
Twist pitch  & $100$ mm \\
Tape width $w_{\mathrm{t}}$ & $10$ mm \\
Characteristic velocity $U_1$, $U_2$ & $0.144$ m/s \\
Hydraulic diameter $D_h$ & $14.3$ mm \\
\hline
\end{tabular}
}

\label{tab:dim}
\end{table}

\begin{table}[t]
\centering
\caption{Physical properties of fluids and solid wall.}
{ \footnotesize
  \renewcommand{\arraystretch}{1}
  \begin{tabular}{lccc}
\hline
Property & Fluid 1 & Fluid 2 & Solid \\
\hline
Density [kg/m$^3$] & $983.2$ & $998.2$ & $2713$ \\
Specific heat [J/kg$\cdot$K] & $4184$ & $4182$ & $917$ \\
Thermal conductivity [W/m$\cdot$K] & $0.654$ & $0.602$ & $202$ \\
Dynamic viscosity [Pa$\cdot$s] & $4.67\times10^{-4}$ & $8.94\times10^{-4}$ & -- \\
\hline
\end{tabular}}
\label{tab:prop}
\end{table}

\begin{figure}[t]
\centering
\includegraphics[width=\columnwidth]{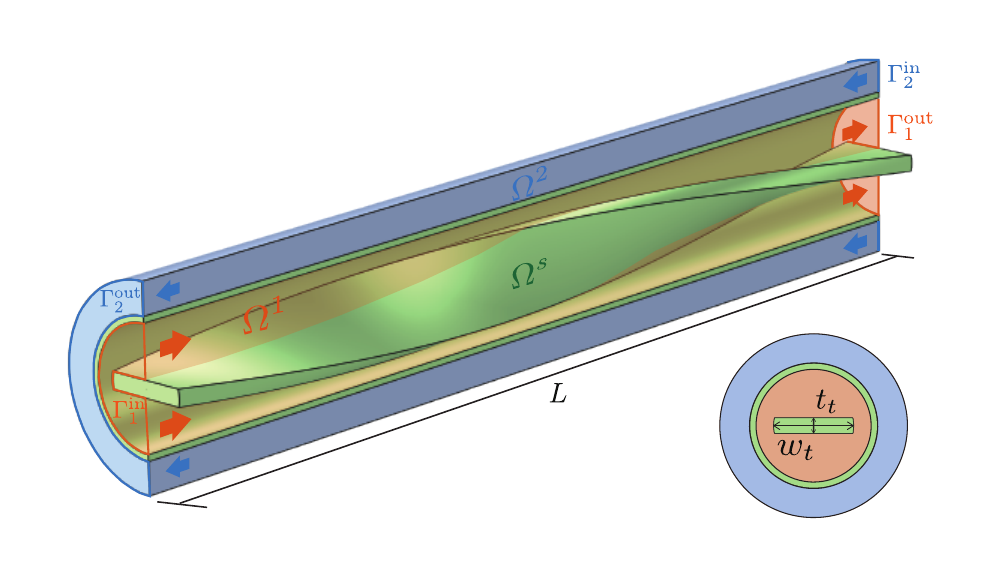}
\caption{Reference design with twisted tape inserted inside the inner pipe.}
\label{fig:setup_tape}
\end{figure}

To demonstrate the effectiveness of the optimized designs, we evaluate the performance of a reference design, whose heat transfer is enhanced by one the most effective performance-enhancement technique, i.e., inserting a twisted tape inside the inner pipe, as illustrated in Fig.~\ref{fig:setup_tape}.
Since the twisted tape is not symmetrical like the straight pipe, we model the entire domain in the HF model for the reference design with twisted tape.
In Fig.~\ref{fig:setup_tape}, the twisted tape and only half regions of the fluid and solid wall are shown for visibility.

\subsection{Numerical implementation of the low-fidelity model}
\label{subsec:impl}
The governing equations Eqs.~(\ref{eq:continuous})--(\ref{eq:bc_2}) are implemented using COMSOL Multiphysics 6.2.
The analysis domain in the LF model is entirely discretized using $\mathbb{P}_1$ tetrahedral finite elements.
The design domain $D$ is discretized with the maximum element size of $0.5$ mm, resulting in a total of 1.8 million elements and 0.31 million design variables.
The initial design variable field $\gamma$ is defined as a uniform field with a value of $0.5$.

\begin{figure}[t]
\centering
\includegraphics[width=\columnwidth]{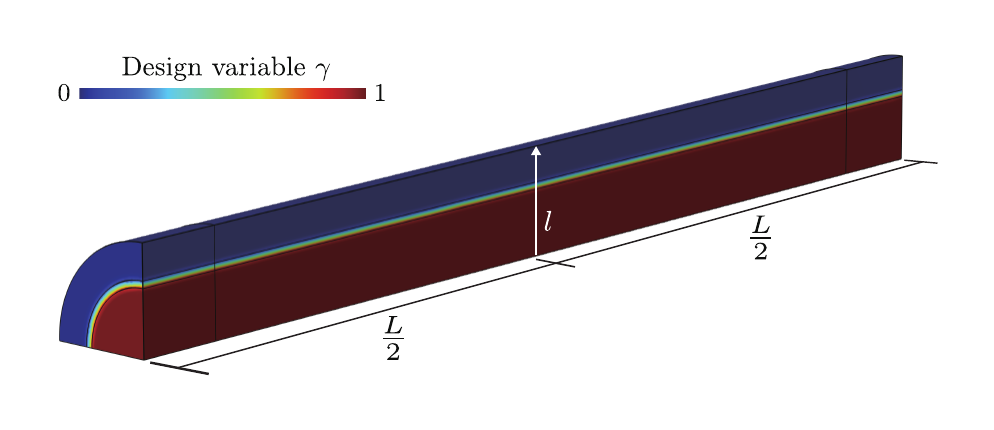}
\caption{Design variable field used for the calibration of the LF model.}
\label{fig:xPhys_LF}
\end{figure}

To ensure that the LF model represents the HF model as accurately as possible, we calibrate the parameters of the LF model for a given design variable field.
Fig.~\ref{fig:xPhys_LF} illustrates the design variable field used for the calibration process, which represents a simple straight pipe structure shown in Fig.~\ref{fig:setup}b.
We perform HF evaluation on this design to obtain reference pressure drop and temperature difference between the inlet and outlet for each fluid.
Subsequently, we calibrate the LF model by fitting the permeability parameters $\kappa^{\max}_i$ of Darcy flow in Eqs.~(\ref{eq:kappa1}) and (\ref{eq:kappa2}) and the sharpness factor $\sigma$ of thermal conductivity interpolation in Eq.~(\ref{eq:ck}), so that the pressure drop and temperature difference of each fluid match those of the HF model.
After several trials of calibration, those parameters are determined for each inlet velocity condition as summarized in Table~\ref{tab:calib}.
The convexity parameters $q_1$ and $q_2$ are set to $900$ to ensure sufficiently low permeabilities in the intermediate region of $\hat{\tilde{\gamma}}$.

\begin{figure}[t]
\centering
\includegraphics[width=\columnwidth]{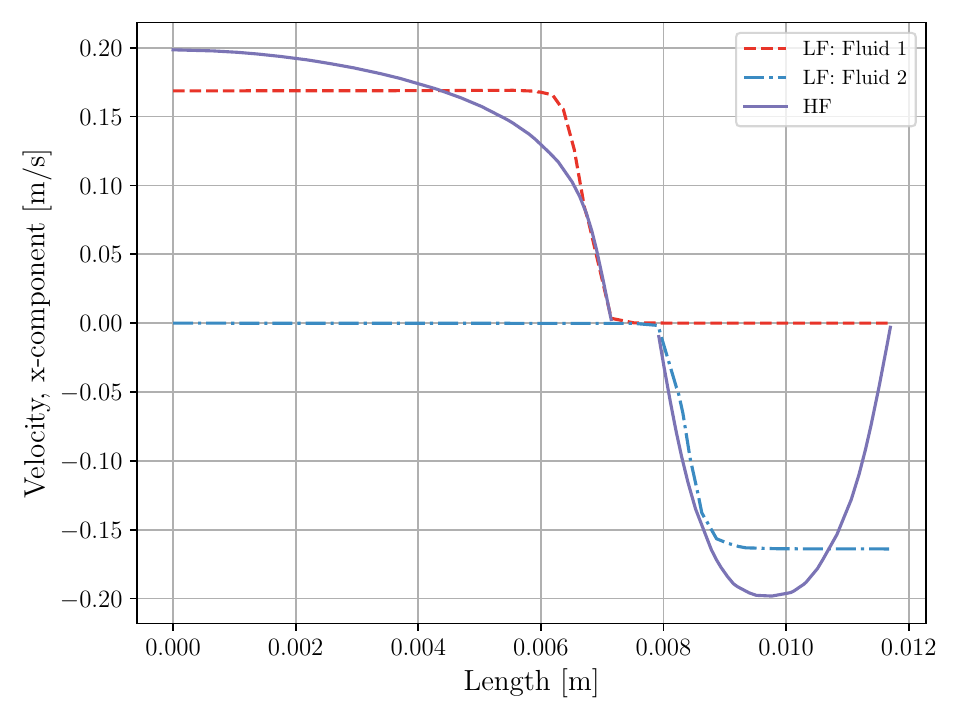}
\caption{Velocity fitting along the evaluation line for $u_{\mathrm{in},1}=u_{\mathrm{in},2}=0.144$ m/s.}
\label{fig:line_u_plot}
\end{figure}

\begin{figure}[t]
\centering
\includegraphics[width=\columnwidth]{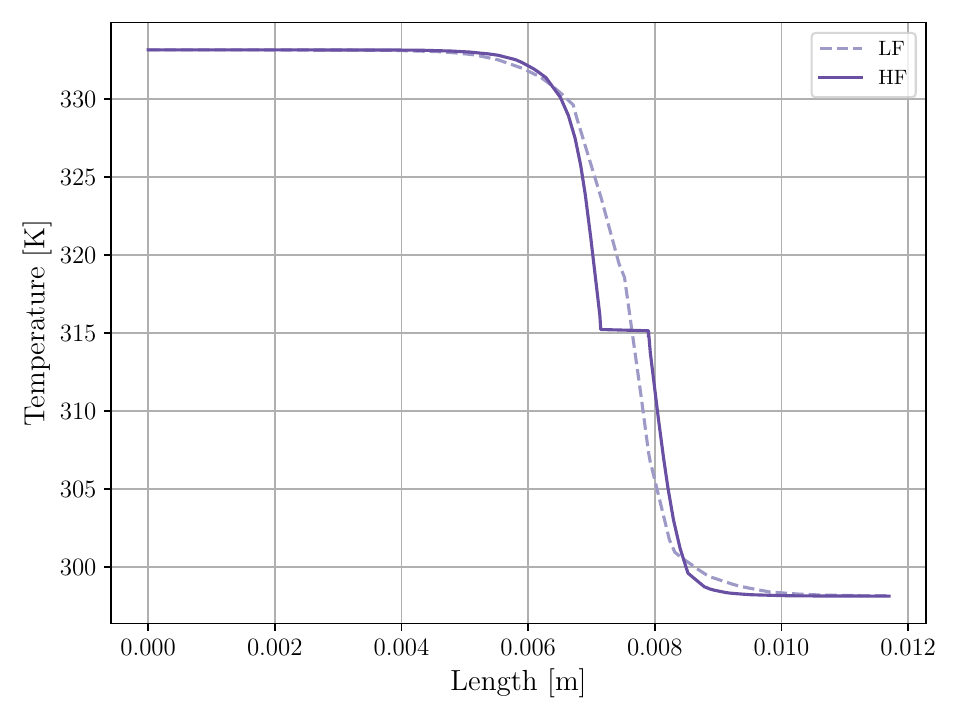}
\caption{Temperature fitting along the evaluation line for $u_{\mathrm{in},1}=u_{\mathrm{in},2}=0.144$ m/s.}
\label{fig:line_temp_plot}
\end{figure}

The fitting examples of the velocity, calculated from the Darcy flow model, and temperature along the evaluation line, visualized as the white line at the center of the design domain in Fig.~\ref{fig:xPhys_LF}, are illustrated in Fig.~\ref{fig:line_u_plot} and Fig.~\ref{fig:line_temp_plot}, respectively.
Fig.~\ref{fig:line_u_plot} shows that each fluid flows only in its respective region, with velocities limited to nearly zero in the opposite fluid region and solid wall region.
This figure also demonstrates a good agreement around the interface between the two fluids and solid wall, but some discrepancies are observed near the center of each fluid region and outer wall.
At the center of each fluid region, the Darcy flow model tends to underestimate the velocity compared to the HF model, because the convexity parameters are set to high values to restrict the flow in the intermediate region.
At the outer wall in the fluid 2 region, the Darcy flow model cannot track the velocity gradient due to the lack of viscous shear terms in the momentum equation.
On the other hand, Fig.~\ref{fig:line_temp_plot} illustrates a good agreement in the temperature distribution between the LF and HF models, except inside the solid wall region.
In this region, the HF model shows a constant temperature distribution due to the high thermal conductivity of aluminum, while the LF model exhibits a temperature gradient due to the Gaussian interpolation function.
We consider that these discrepancies in the LF model are acceptable for the TO process, as the overall fitting of pressure drop and temperature difference is well achieved through the calibration.

\begin{table}[t]
\centering
\caption{Calibration parameters of the LF model for each inlet velocity condition.}
{ \footnotesize
  \renewcommand{\arraystretch}{1}
  \begin{tabular}{lccc}
\hline
($u_{\mathrm{in},1}, u_{\mathrm{in},2}$) [m/s] & $\kappa^{\max}_1$ & $\kappa^{\max}_2$ & $\sigma_\text{cl}$ \\
\hline
($0.144$, $0.144$) & $1.69\times10^{-2}$ & $7.6\times10^{-3}$ & $1.82\times10^{-2}$ \\
($0.217$, $0.144$) & $1.3\times10^{-2}$ & $7.6\times10^{-3}$ & $1.76\times10^{-2}$ \\
($0.289$, $0.144$) & $1.07\times10^{-2}$ & $7.6\times10^{-3}$ & $1.88\times10^{-2}$ \\
($0.361$, $0.144$) & $9.15\times10^{-3}$ & $7.6\times10^{-3}$ & $1.9\times10^{-2}$ \\
($0.433$, $0.144$) & $8.05\times10^{-3}$ & $7.6\times10^{-3}$ & $1.94\times10^{-2}$ \\
\hline
\end{tabular}}
\label{tab:calib}
\end{table}

To solve the optimization problem in Eq.~(\ref{eq:optimization_problem_total}), the design sensitivity of the objective function $J$ with respect to the original design variables $\gamma$ is calculated using the chain rule, where the adjoint method is employed to compute the sensitivity of $J$ with respect to the projected design variable field $\hat{\tilde{\gamma}}$.
The method of moving asymptotes (MMA)~\cite{svanberg1987method} is employed as the optimization algorithm to update the design variable field based on the computed sensitivities.

\subsection{Numerical implementation of the high-fidelity model}
\label{subsec:val_hig}

\begin{figure}[t]
\centering
\includegraphics[width=\columnwidth]{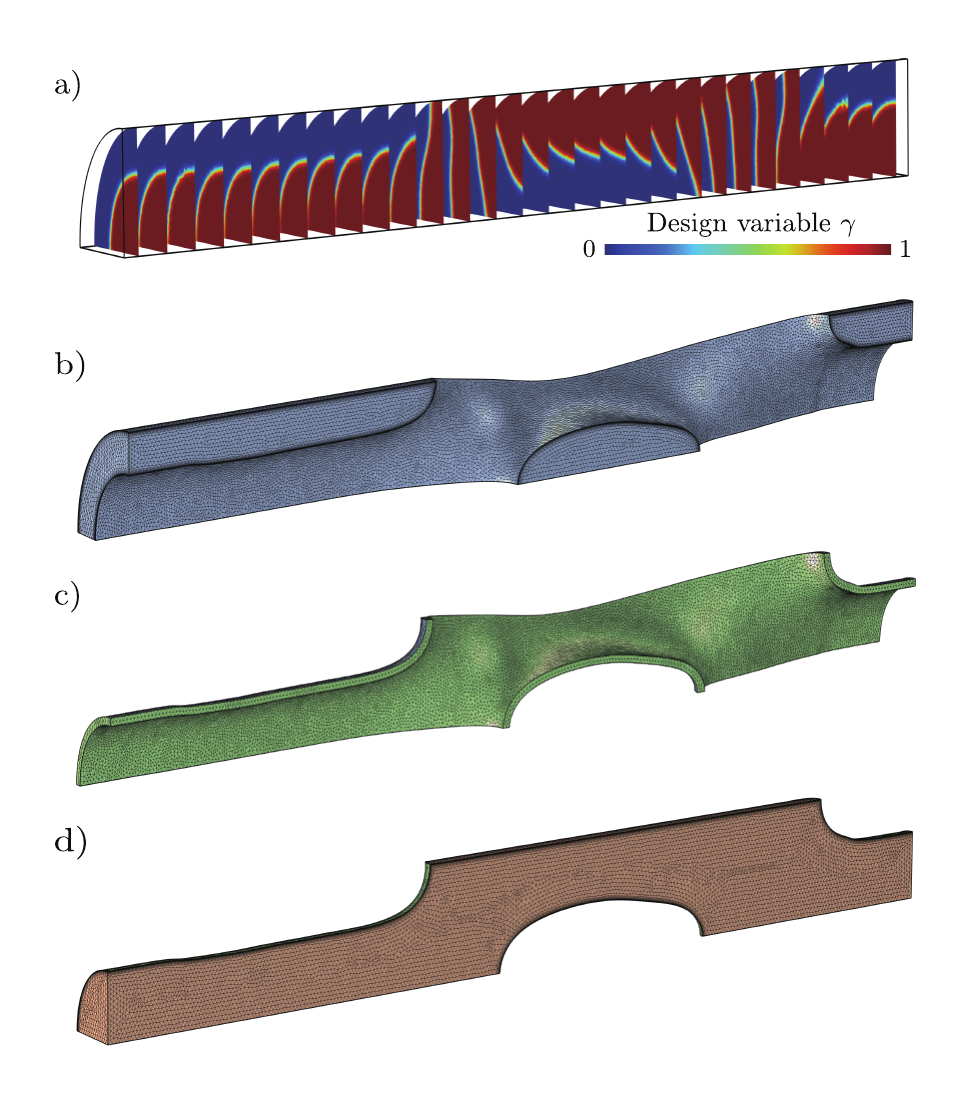}
\caption{Mesh examples of the HF model: a) design variable field; b) fluid 1 mesh; c) solid wall mesh; d) fluid 2 mesh.}
\label{fig:hf_mesh}
\end{figure}

The analysis domain in the HF model is discretized using body-fitted meshes for each subdomain, as illustrated in Fig.~\ref{fig:hf_mesh}.
The body-fitted meshes are constructed by calculating the marching cubes of the design variable field $\hat{\tilde{\gamma}}$ to extract isosurfaces that define the interfaces between the fluid and solid wall regions, and then applying a meshing algorithm to each region, using built-in functions in the COMSOL.
The lower and upper thresholds for defining the solid wall region are set to $\gamma^l=0.42$ and $\gamma^u=0.58$, so that the solid wall region of straight wall design has the same volume as that in Fig.~\ref{fig:setup}b).
The meshing process is automated with in-house MATLAB scripts to be consistent for all optimized designs.

The meshes consist of tetrahedral elements for the bulk region and prismatic elements for the boundary layer region along all walls in contact with the fluids.
The wall function is applied to the fluid-solid interfaces to reduce the computational cost for resolving the near-wall turbulent flow.
This special treatment is very difficult to implement in the density-based TO, justifying the use of the proposed multifidelity approach that leverages the separate modeling of the optimization and evaluation processes.

\begin{figure*}[t]
\centering
\includegraphics[width=0.9\textwidth]{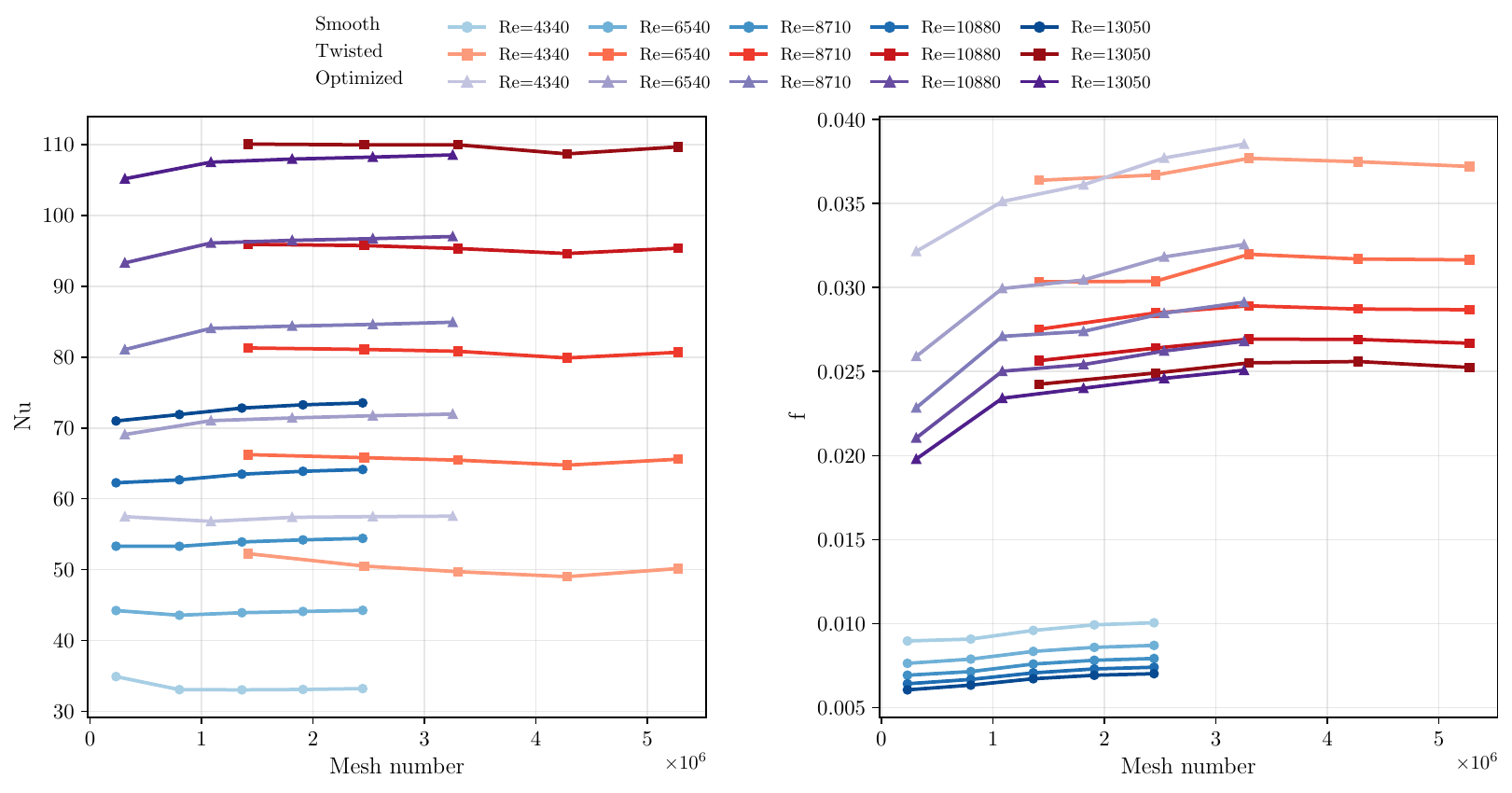}
\caption{Mesh dependency of the HF model regarding Nusselt number Nu and friction factor f.}
\label{fig:mesh_dependency}
\end{figure*}

The maximum element size and the number of prismatic meshes are determined based on the mesh dependency verification, as shown in Fig.~\ref{fig:mesh_dependency}.
The combinations of these two parameters investigated in this verification are summarized in Table~\ref{tab:mesh_setting}.
Smooth and optimized models have the same mesh settings, while the twisted model has different mesh settings because it is constructed as the full domain, unlike the other two models that are constructed as the quarter domain.
For optimized model, a representative optimized design obtained from the LF optimization process is used for the mesh dependency verification, i.e., the case with $\alpha=1.0$ and $\text{Re}=13{,}000$.
Each mesh setting is applied to all five inlet velocity conditions to confirm the mesh dependency for a wide range of turbulent flow conditions.
In this verification, we evaluate the Nusselt number Nu and friction factor f of fluid 1 for each mesh setting, defined as follows:
\begin{align}
  \text{Nu} & = \frac{h D_h}{k_1}, \\
  \text{f} & = \frac{\Delta p_1 D_h}{2 \rho_1 u_{\mathrm{in}, 1}^2 L},
\end{align}
\noindent where $h$ is the convective heat transfer coefficient calculated from the HF simulation results; $D_h = 2R_2$ is the hydraulic diameter; and $\Delta p_1$ is the pressure drop of fluid 1.
Fig.~\ref{fig:mesh_dependency} shows that Nu and f converge as the mesh is refined, and the results for the second most refined case are close to those for the most refined case, indicating that the second most refined case is sufficient for accurate evaluation while reducing computational costs.
To ensure the convergence of the mesh dependency, we also calculate the relative errors of Nu and f for the second most refined cases compared to the most refined cases.
The average relative errors across the five inlet velocity conditions are $0.4\%$ for Nu and $1.3\%$ for f in the smooth model; $1.3\%$ for Nu and $0.6\%$ for f in the twisted model; $0.3\%$ for Nu and $2.2\%$ for f in the optimized model.
These small errors confirm the convergence of the mesh dependency.
Therefore, we employ these mesh settings for each model throughout this study.

\begin{table}[t]
\centering
\caption{Mesh settings for each model and case.}
\label{tab:mesh_setting}
{\footnotesize
\renewcommand{\arraystretch}{1.15}
\begin{tabular}{llccccc}
\hline
Model & Parameter & 1 & 2 & 3 & 4 & 5\\
\hline
\multirow{2}{*}{Smooth}
& Boundary layer              & 5       & 8       & 9       & 10    & 11 \\
& Max mesh size & 1/24    & 1/36    & 1/43    & 1/48    & 1/52 \\
\hline
\multirow{2}{*}{Twisted}
& Boundary layer              & 5       & 6       & 7       & 8       & 9 \\
& Max mesh size & 1/24    & 1/30    & 1/33    & 1/36    & 1/39 \\
\hline
\multirow{2}{*}{Optimized}
& Boundary layer              & 5       & 8       & 9       & 10    & 11 \\
& Max mesh size & 1/24    & 1/36    & 1/43    & 1/48    & 1/52 \\
\hline
\end{tabular}
}
\end{table}

\begin{figure}[t]
\centering
\includegraphics[width=\columnwidth]{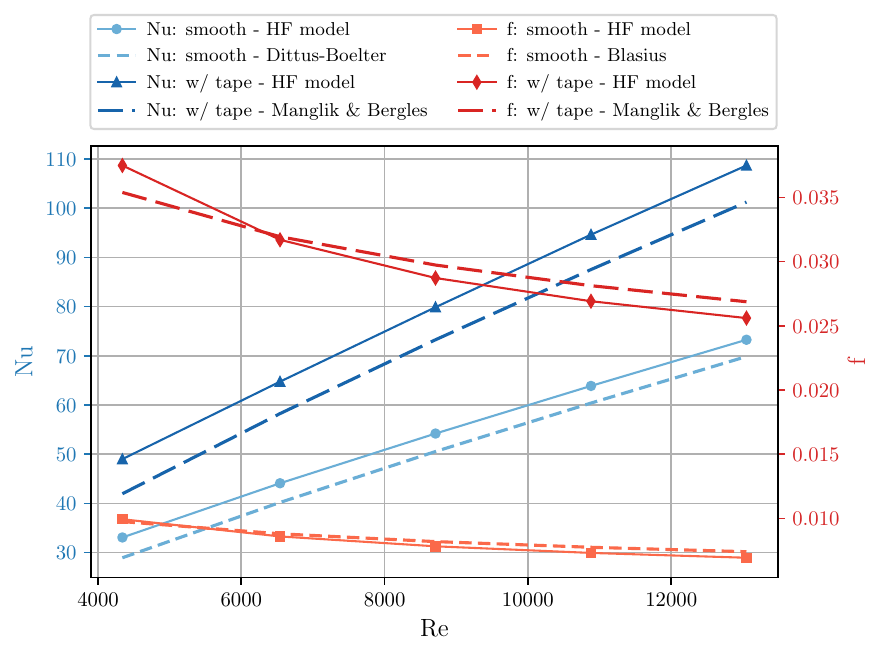}
\caption{Comparison of Nusselt numbers Nu and friction factors f between the HF model and correlations; Nu correlation: Dittus-Boelter equation for smooth pipe and Manglik and Bergles equation~\cite{manglik1993heat} for twisted tape insertion; f correlation: Blasius equation for smooth pipe and Manglik and Bergles equation~\cite{manglik1993heat} for twisted tape insertion.}
\label{fig:validation}
\end{figure}

The governing equations Eqs.~(\ref{eq:RANS})--(\ref{eq:tm_ch}) of the fluid flow and heat transfer are implemented using the COMSOL.
To validate the implemented HF model, we compare the Nu and f obtained from the HF model with those calculated using widely used empirical correlations for turbulent flow in smooth pipes, i.e., the Dittus-Boelter equation for Nu and the Blasius equation for f, defined as follows:
\begin{equation}
\begin{aligned}
\text{Nu} & = 0.023 \text{Re}^{0.8} \text{Pr}^{0.4}, \\
\text{f} & = 0.0791 \text{Re}^{-0.25},
\end{aligned}
\label{eq:smooth_correlation}
\end{equation}
\noindent
where Re ranges from $4{,}300$ to $13{,}000$ in this study, and Pr is constant at 3.0 for fluid 1, which is calculated as $\text{Pr} = \mu_1 {c_\text{p}}_1 / k_1$.
The same validation is also performed for the reference design with twisted tape, using the empirical correlations proposed by Manglik and Bergles~\cite{manglik1993heat} for both Nu and f, defined as follows:
\begin{equation}
\begin{aligned}
\text{Nu}_\infty & = 0.023 \text{Re}^{0.8} \text{Pr}^{0.4} \left(\frac{\pi}{\pi-4 t_t / D_h}\right)^{0.8} \left( \frac{\pi+2-2t_t/D_h}{\pi-4t_t/D_h} \right)^{0.2}, \\
\text{Nu} & = \text{Nu}_\infty \left( 1+\frac{0.769}{L/t_t} \right) , \\
\text{f}_\infty & = \frac{0.0791}{\text{Re}^{0.25}} \left(\frac{\pi}{\pi-4 t_t / D_h}\right)^{1.75} \left( \frac{\pi+2-2t_t/D_h}{\pi-4t_t/D_h} \right)^{1.25}, \\
\text{f} & = \text{f}_\infty \left( 1+ \frac{2.752}{(L/t_t)^{1.29}} \right).
\end{aligned}
\label{eq:twisted_correlation}
\end{equation}
Fig.~\ref{fig:validation} illustrates the comparison results for the smooth and twisted models, showing a good agreement between the HF model and the empirical correlations for both Nu and f across the five Reynolds numbers.
Additionally, the relative errors of Nu and f obtained from the HF model compared to those calculated using the empirical correlations are calculated for each Reynolds number.
The average relative errors across the five Reynolds numbers are $8.3\%$ for Nu and $4.1\%$ for f in the smooth model; $10\%$ for Nu and $3.8\%$ for f in the twisted model.
These errors confirm the validity of the HF model implemented in the COMSOL for simulating turbulent heat transfer in the DPHX model.

The key modeling assumptions for the HF model are as follows: $\mathbb{P}_1$ finite elements are used for all variables; steady state with incompressible single phase flow of Newtonian fluid is assumed; physical properties are constant; gravity/buoyancy, viscous dissipation, and radiation heat transfer are neglected; no slip boundary conditions are applied at boundaries except inlets, outlets, and symmetry planes.
The governing equations are solved using a stationary segregated solver, where the linear systems in each segregated block are solved using GMRES with algebraic multigrid (AMG) preconditioning, with consistent solver settings for all cases to ensure a fair comparison among the cases.

\subsection{Optimized design}
\label{subsec:optimized}
Using the calibrated LF model described in section~\ref{subsec:impl}, we perform TO for the DPHX model illustrated in Fig.~\ref{fig:setup}a.
The optimization settings are summarized in Table~\ref{tab:opt_settings}.
In order to promote stable calculations at the early stage of the optimization process and obtain clear designs with distinct fluid and solid wall regions at the final stage, we employ a continuation approach for several parameters during the optimization process, as summarized in Table~\ref{tab:continuation}.
The thermal conductivity sharpness $\sigma$ in Eq.~(\ref{eq:ck}) is gradually decreased to the calibrated value to avoid convergence issues due to the steep change in thermal conductivity at the fluid-solid interface, calculated as multiplying the calibrated value $\sigma_\text{cl}$ by the continuation factor $\sigma_\text{ct}$ listed in Table~\ref{tab:continuation}, i.e., $\sigma=\sigma_\text{cl} \sigma_\text{ct}$.

The trade-off parameter $w_p$ in Eq.~(\ref{eq:optimization_problem_total}) is set as $w_p=\alpha w_\text{ct}$, where $w_\text{ct}$ is increased by multiplying $10$ every $50$ iterations to gradually enforce the pressure drop constraints, while $\alpha$ is a scaling factor to adjust the trade-off parameter.
The penaly parameter $w_p$ essentially determines the trade-off between maximizing the heat transfer and satisfying the pressure drop constraints.
Therefore, we perform optimizations for five different values of $\alpha$ as seeding paramters in the multifidelity approach, to explore various trade-offs and obtain diverse optimized designs.
Additionally, we explore five different inlet velocities for fluid 1 also as seeding parameters, while keeping the inlet velocity for fluid 2 constant, to investigate the effect of Reynolds number on the optimized designs.
In total, we conduct $25$ optimization cases by combining the different values of $\alpha$ and inlet velocities for fluid 1 $u_{\mathrm{in},1}$.

\begin{table}[t]
\centering
\caption{Optimization settings for the LF model.}
{ \footnotesize
  \renewcommand{\arraystretch}{1}
  \begin{tabular}{lc}
\hline
Parameter & Value \\
\hline
Maximum element size & $0.021$ \\
Filter radius $r_\text{min}$ & $0.167$ \\
Initial design variable field & $0.5$ \\
Projection threshold $\eta$ & $0.5$ \\
Scaling factor of trade-off parameter $\alpha$ & $0.2, 0.4, 0.6, 0.8, 1.0$ \\
Inlet velocity fluid 1 $u_{\mathrm{in},1}$ & $0.144, 0.217, 0.289, 0.361, 0.433$ \\
Inlet velocity fluid 2 $u_{\mathrm{in},2}$ & $0.144$ \\
Pressure drop constraints: fluid 1 $\Delta p_1^0$ & $3.40, 6.67, 10.8, 15.7, 21.4$ \\
Pressure drop constraints: fluid 2 $\Delta p_2^0$ & $14.3$ \\
Maximum MMA iterations & $300$ \\
\hline
\end{tabular}}
\label{tab:opt_settings}
\end{table}

\begin{table}[t]
\centering
\caption{Continuation settings for the LF model.}
{ \footnotesize
  \renewcommand{\arraystretch}{1}
  \setlength{\tabcolsep}{3pt} 
  \begin{tabular}{lcccccc}
\hline
 & 1st & 2nd & 3rd & 4th & 5th & 6th \\
\hline
Max iteration & $50$ & $100$ & $150$ & $200$ & $250$ & $300$ \\
Projection sharpness $\beta$ & $4$ & $8$ & $16$ & $16$ & $16$ & $16$ \\
Trade-off parameter $w_\text{ct}$ & $10^{-2}$ & $10^{-1}$ & $10^{0}$ & $10^{1}$ & $10^{2}$ & $10^{3}$ \\
Thermal conductivity sharpness $\sigma_\text{ct}$ & $2^2$ & $2^1$ & $1$ & $1$ & $1$ & $1$ \\
\hline
\end{tabular}}
\label{tab:continuation}
\end{table}

\subsubsection{Low-fidelity optimization results}
\label{subsubsec:lf_opt}

\begin{figure}[t]
\centering
\includegraphics[width=\columnwidth]{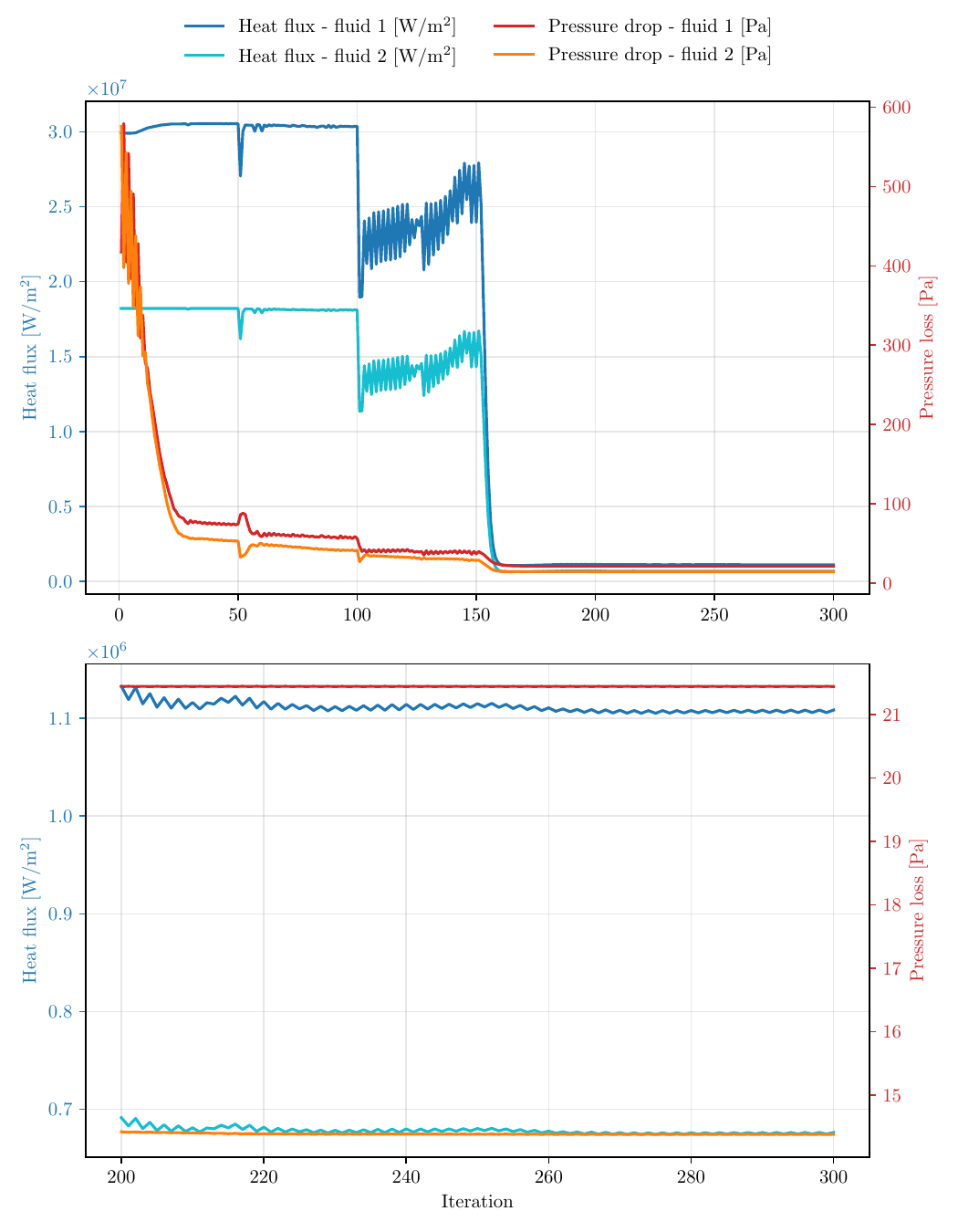}
\caption{Objective function history during the optimization process for $u_{\mathrm{in},1}=0.433$ m/s and $\alpha=1.0$; (top) the entire history; (bottom) the history from iteration 200 to 300.}
\label{fig:optimization_history}
\end{figure}

\begin{figure*}[t]
\centering
\includegraphics[width=\textwidth]{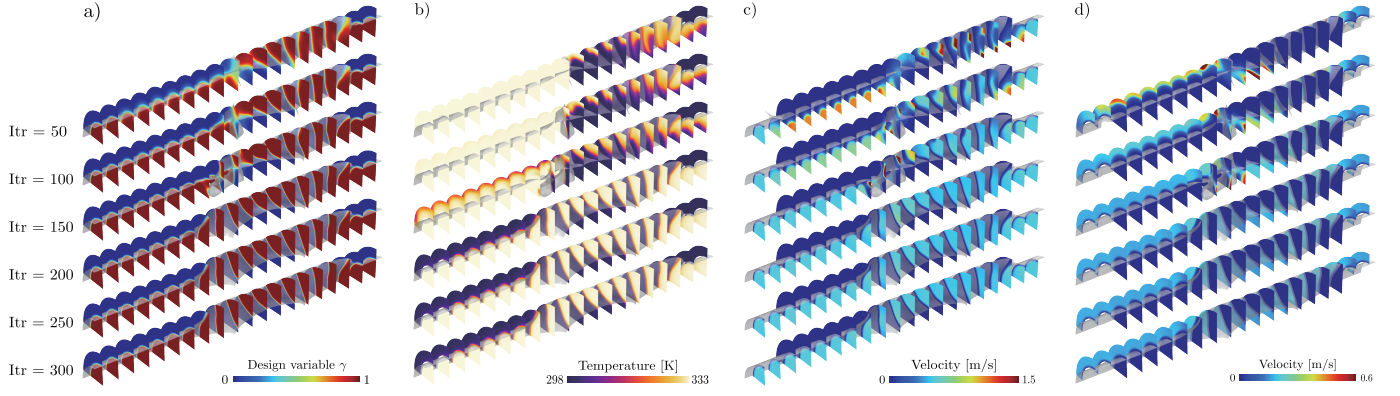}
\caption{History of the a) design variable field, b) temperature distribution, c) velocity distribution for fluid 1, and d) velocity distribution for fluid 2 at every 50 iterations during the optimization process for $u_{\mathrm{in},1}=0.433$ m/s and $\alpha=1.0$.}
\label{fig:iterations}
\end{figure*}

Fig.~\ref{fig:optimization_history} illustrates the objective function history during the optimization process for a case with $u_{\mathrm{in},1}=0.433$ m/s and $\alpha=1.0$, where the entire history and the history from iteration 200 to 300 are shown in the top and bottom plots, respectively.
The $J_i^h$ are dimensionalized to be dimesional heat fluxes, while $\Delta p_i$ are dimesionalized to be dimensional pressure drops.
Fig.~\ref{fig:iterations} shows the history of the design variable field, temperature distribution, and velocity distribution at every 50 iterations during the optimization process for the same case.
The dimensionalized temperature and velocity fields are visualized as slices vertically cutting through the center of the pipes, and the center surfaces of the solid wall region are visualized as isosurfaces of the design variable field at $\hat{\tilde{\gamma}}=0.5$.

At the early iterations, the heat flux is much higher than the converged value as shown in Fig.~\ref{fig:optimization_history}, which is consistent with the temperature distributions shown in Fig.~\ref{fig:iterations}b.
This is because the design variable field at the early iterations, shown in Fig.~\ref{fig:iterations}a, contains large intermediate regions with high thermal conductivity and permeability, resulting in a large amount of heat transfer between the two fluids.
At the transition from the first to the second continuation stages at iteration 50 and from the second to the third continuation stages at iteration 100, the heat flux slightly decreased due to the increase in the projection sharpness $\beta$, which sharpens the design variable field and suppresses the heat transfer in the intermediate regions.
After these transitions, the heat flux gradually increased again as the optimization proceeded in each continuation stage.
In the third continuation stage, due to the increase in the trade-off parameter $w_p$ in addition to the projection sharpness $\beta$, the optimization process started to focus more on reducing the pressure drops by shrinking the intermediate regions, yet still improving the heat flux to some extent by streamling the flow paths.
The search for such trade-off balancing heat transfer and pressure drops caused a large fluctuation in the heat flux during this stage.
Once the optimization entered the fourth continuation stage at iteration 150, the heat flux drastically decreased as the trade-off parameter was further increased, that is, the pressure drop was more strongly weighted in the objective function.
This is because the further increased trade-off parameter forced the design variable field to streamline the flow paths more aggressively, sacrificing the heat transfer performance.
Simultaneously, as the optimization progresses, the pressure drops for both fluids decrease as shown in Fig.~\ref{fig:optimization_history}, as intended by the continuation of the trade-off parameter $w_p$.
As the pressure drop constraints are enforced, the greyscale regions in the design variable field are eliminated, since these regions cause high pressure drops due to their low permeability.
As the flow paths become simpler and more streamlined, the velocity distributions for both fluids also change significantly, as shown in Fig.~\ref{fig:iterations}c and d, additionally contributing to the reduction of pressure drops.
After approximately 250 iterations, in this case, the optimization converges, with the design variable field forming clear fluid and solid wall regions, and the temperature and velocity distributions stabilizing.

\begin{figure*}[t]
\centering
\includegraphics[width=\textwidth]{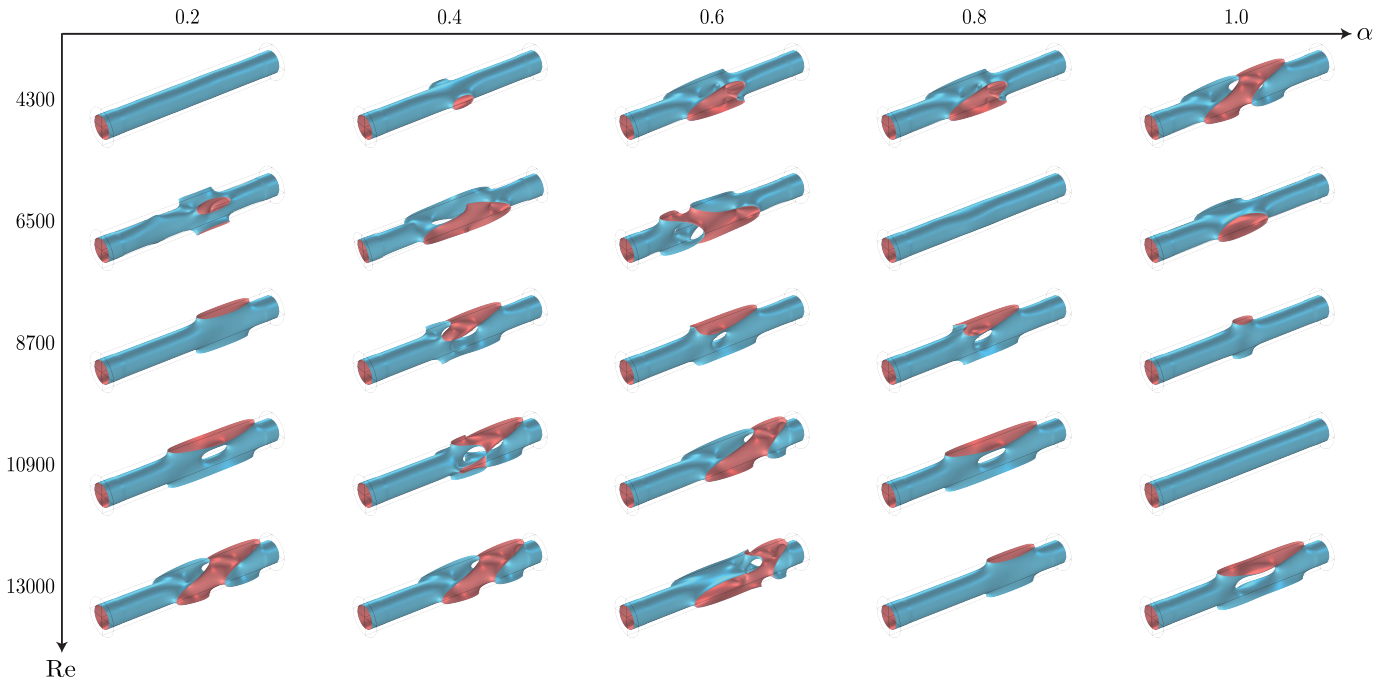}
\caption{Optimized designs obtained from the LF model for all combinations of the inlet velocity $u_{\mathrm{in},1}$ (shown as corresponding Reynolds numbers) and the scaling factor of trade-off parameter $\alpha$; the red and blue surfaces represent the solid wall shapes on the fluid 1 and fluid 2 sides, respectively.}
\label{fig:opt_wall_images}
\end{figure*}

Fig.~\ref{fig:opt_wall_images} illustrates the optimized designs obtained from the LF model for all combinations of $u_{\mathrm{in},1}$ and $\alpha$, where the red and blue surfaces represent the solid wall shapes on the fluid 1 and fluid 2 sides, respectively.
This figure shows that various complex wall structures are obtained depending on the optimization settings.
There is no clear trend in the wall structures with respect to $u_{\mathrm{in},1}$ and $\alpha$, indicating that the optimization results are significantly influenced by the continuation approach and the optimization path taken during the process.
This is confirmed by the fact that some cases result in straight pipe-like structures, as a result of focusing on minimizing pressure drops early in the optimization, which is observed across different optimization settings, e.g., the cases with $u_{\mathrm{in},1}=0.144$ m/s and $\alpha=0.2$ and with $u_{\mathrm{in},1}=0.361$ m/s and $\alpha=1.0$.

\begin{figure}[t]
\centering
\includegraphics[width=\columnwidth]{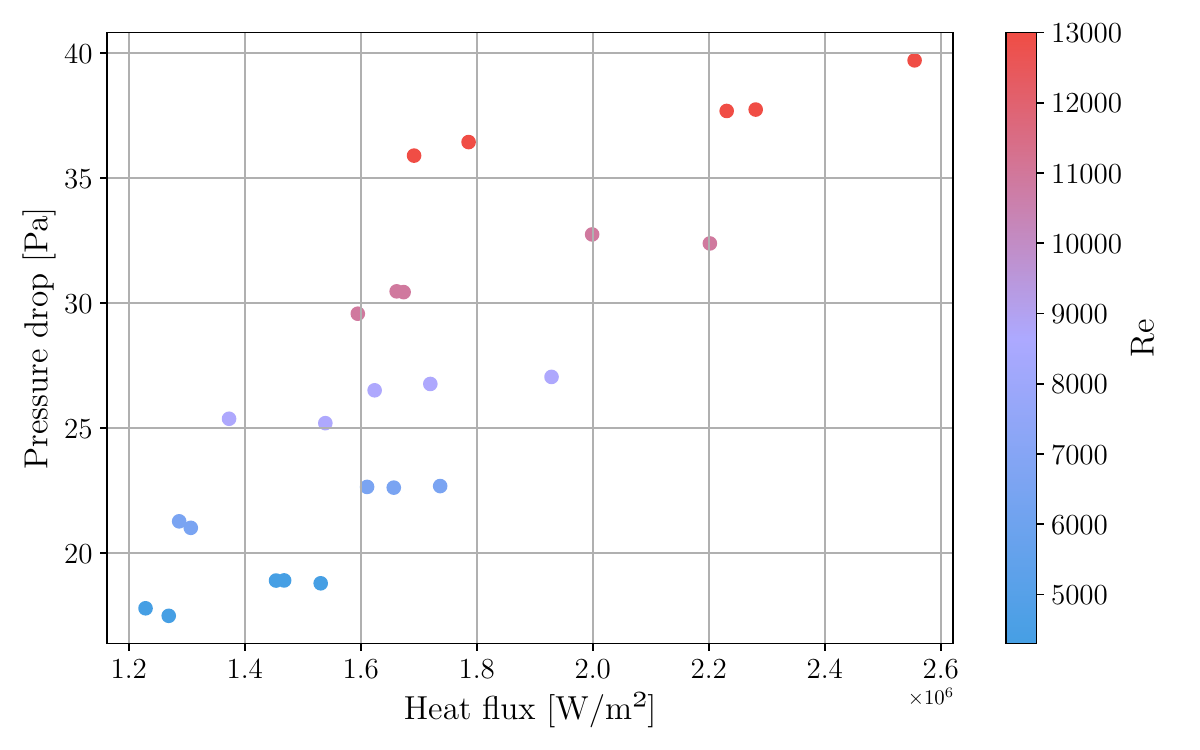}
\caption{Performance comparison of the optimized designs obtained from the LF model.}
\label{fig:LF_obj}
\end{figure}

Fig.~\ref{fig:LF_obj} compares the performance of the optimized designs obtained from the LF model for all optimization cases, where the total values of two fluids are plotted in terms of heat flux and pressure drops.
Note that the dimensional pressure drops and heat fluxes in this figure are calculated by summing up the non-dimensional pressure drops $\Delta p_i$ and heat fluxes $J_i^h$, respectively, and then dimensionalizing them as described in~\ref{ap:dimensionless} and \ref{ap:heat_flux}.
This figure shows a clear trade-off relationship between the heat flux and pressure drops, indicating that higher heat fluxes can be achieved at the cost of increased pressure drops.
Additionally, the results show that higher inlet velocities for fluid 1 lead to higher heat fluxes and pressure drops, which is consistent with the physical understanding of convective heat transfer and fluid flow.

\subsubsection{High-fidelity evaluation results}
\label{subsubsec:hf_eval}

To discuss the performance of the optimized designs, we utilize overall performance metrics, i.e., the total pressure drops, overall heat transfer coefficient, and performance evaluation criteria (PEC), defined as follows:
\begin{align}
  \Delta p_\text{total} & = \Delta p_1 + \Delta p_2, \\
  U & = \frac{q}{\Delta T_\text{lm}}, \\
  \text{PEC} & = \frac{U/U_0}{(\Delta p_\text{total}/\Delta p_\text{total,0})^{1/3}},
\end{align}
\noindent where $q$ is the heat flux from fluid 1 to solid wall region, obtained from the HF model; the subscript $0$ denotes the values for the smooth pipe reference design; $\Delta T_\text{lm}$ is the log-mean temperature difference between the two fluids, calculated as
\begin{align}
  \Delta T_\text{lm} = \frac{(T_{\mathrm{in},1} - T_{\mathrm{out},2}) - (T_{\mathrm{out},1} - T_{\mathrm{in},2})}{\ln\left(\frac{T_{\mathrm{in},1} - T_{\mathrm{out},2}}{T_{\mathrm{out},1} - T_{\mathrm{in},2}}\right)}.
\end{align}

\begin{figure}[t]
\centering
\includegraphics[width=\columnwidth]{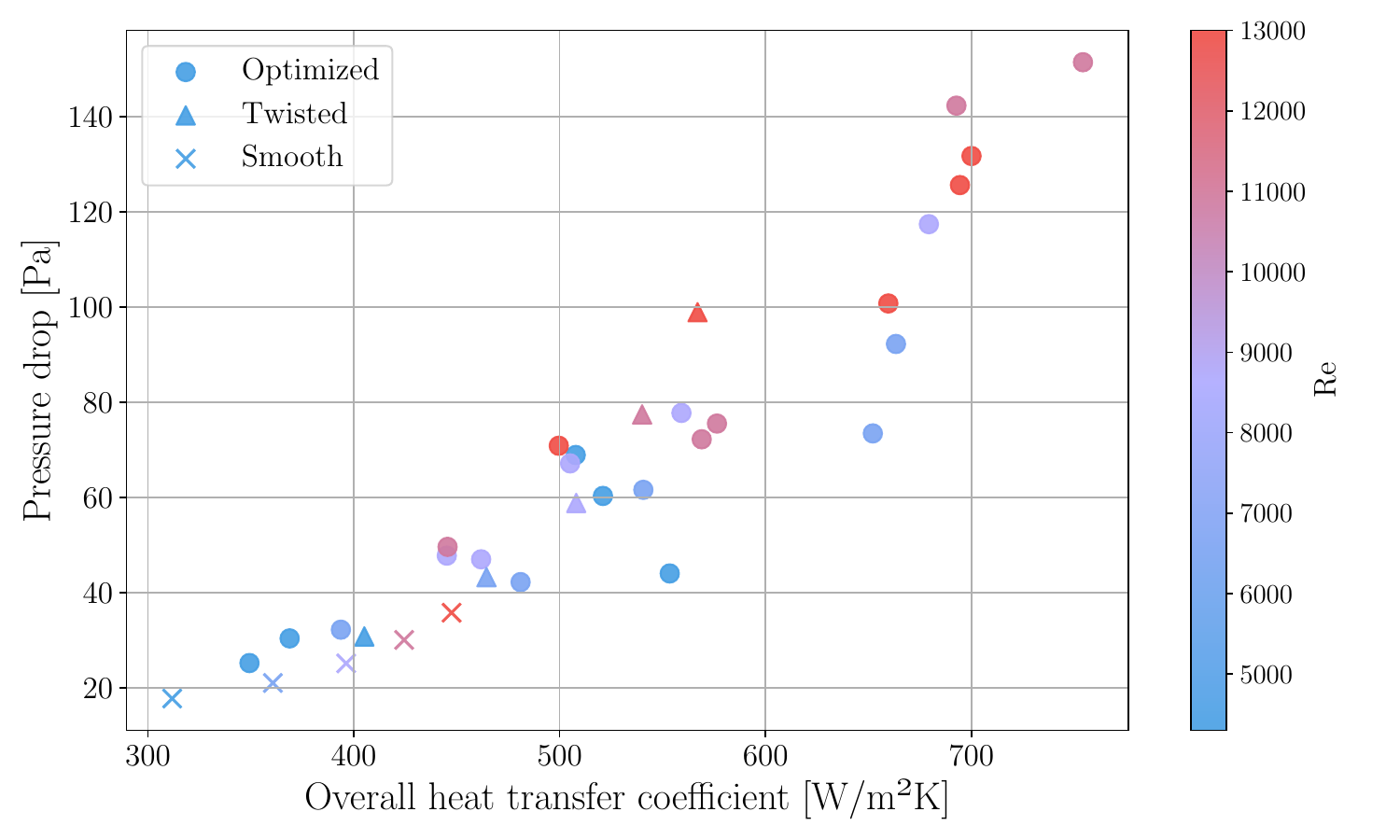}
\caption{Performance comparison of the high-fidelity evaluation results between the optimized designs and reference designs.}
\label{fig:performance_HF}
\end{figure}

Fig.~\ref{fig:performance_HF} compares the pressure drops and overall heat transfer coefficients between the optimized designs and reference designs, evaluated using the HF model, where the Reynolds number for fluid 1 is colored.
Similar to the LF model results shown in Fig.~\ref{fig:LF_obj}, the HF model results exhibit a trade-off relationship between the overall heat transfer coefficient and pressure drops for the optimized designs, with higher Reynolds numbers leading to higher overall heat transfer coefficients and pressure drops.
For reference designs, twisted tape insertion significantly increases both the overall heat transfer coefficient and pressure drops compared to the smooth pipe design, demonstrating its effectiveness as a heat transfer enhancement technique.
However, especially at high Reynolds numbers, the twisted tape insertion results in excessively high pressure drops, which makes the technique less efficient in terms of PEC.
On the other hand, the optimized designs obtained from the LF model exhibit a wide range of performance, with some designs achieving higher overall heat transfer coefficients than the reference designs while maintaining pressure drops comparable to those of the twisted tape insertion.
This trend is more likely to be observed at higher Reynolds numbers, indicating that the TO approach using the Darcy flow-based LF model can effectively explore high-performance designs that balance heat transfer enhancement and pressure drop mitigation.

When comparing the LF model results in Fig.~\ref{fig:LF_obj} with the HF model results in Fig.~\ref{fig:performance_HF}, the LF model tends to underestimate the pressure drops for optimized designs.
This is because the Darcy flow model used in the LF model does not account for viscous shear effects and turbulence-induced momentum losses, leading to lower predicted pressure drops compared to the HF model that incorporates these effects through the RANS equations.
More detailed discussions on the differences between the LF and HF models are provided in section~\ref{subsubsec:diff}.

\begin{figure*}[t]
\centering
\includegraphics[width=\textwidth]{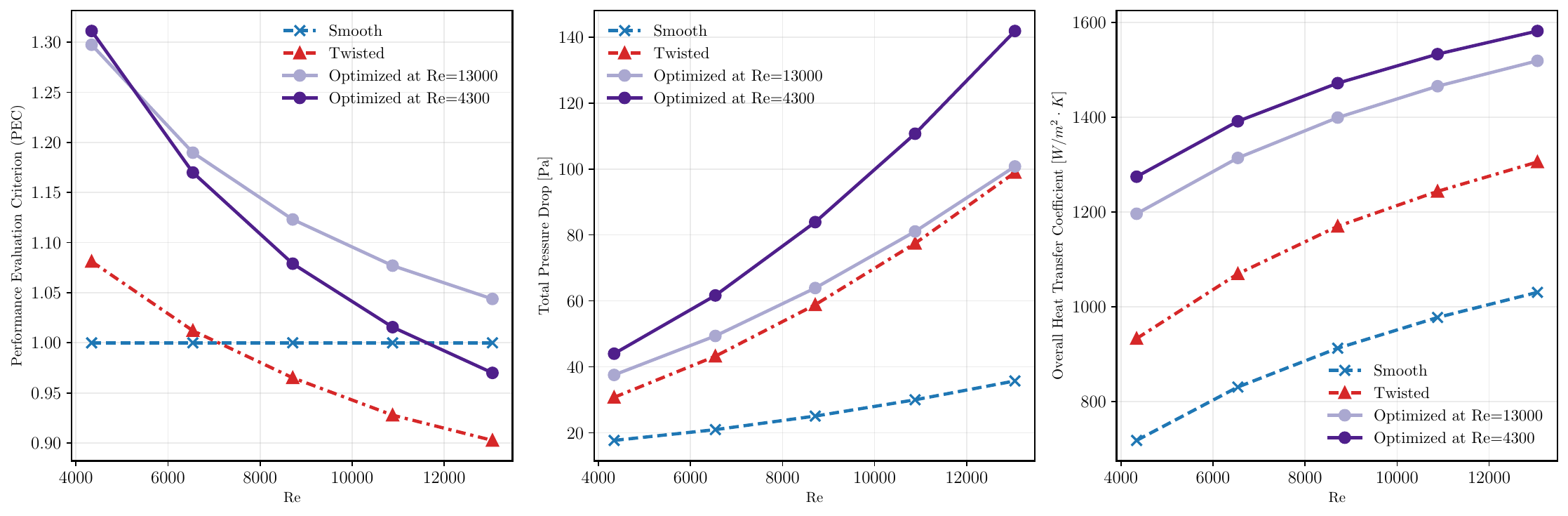}
\caption{Performance comparison between the optimized designs and reference design.}
\label{fig:PEC}
\end{figure*}

To further evaluate the performance of the optimized designs, we select the designs that achieved the highest PEC at the lowest and highest Reynolds numbers for fluid 1, i.e., $\text{Re}=4{,}300$ and $13{,}000$, and run static simulations using the HF model at all five Reynolds numbers for each selected design.
Fig.\ref{fig:PEC} compares the values of PEC, total pressure drops, and overall heat transfer coefficients between the optimized designs and reference designs.
Twisted tape insertion improves overall heat transfer coefficients $26.7$\%--$30.0$\% compared to the smooth pipe design, but results in a significant increase in total pressure drops by $73.7$\%--$177$\%, leading to PEC values lower than $1.0$ at three high Reynolds numbers of $\text{Re} > 6{,}500$.
On the other hand, the optimized design at $\text{Re}=13{,}000$ achieves overall heat transfer coefficients improvements of $47.4$\%--$66.7$\%, compared to the smooth pipe design, while maintaining total pressure drops increases of $112$\%--$182$\%, which is comparable to those of the twisted tape insertion.
As a result, the optimized design at $\text{Re}=13{,}000$ achieves PEC values higher than $1.0$ at all Reynolds numbers, ranging from $1.04$ to $1.30$.
The optimized designs at $\text{Re}=4{,}300$ achieve overall heat transfer coefficients improvements of $53.5$\%--$77.6$\%, which are even higher than those of the optimized designs at $\text{Re}=13{,}000$, but at the cost of significantly increased total pressure drops by $148$\%--$297$\%, compared to the smooth pipe design.
Consequently, the optimized design at $\text{Re}=4{,}300$ attains PEC values greater than $1.0$ at every Reynolds number except for $\text{Re}=13{,}000$, with PEC values ranging from $0.97$ to $1.31$.
This result indicates that the optimized designs with high PEC values at higher Reynolds numbers tend to maintain high PEC values across a wide range of Reynolds numbers, demonstrating their robustness and effectiveness in enhancing heat transfer while managing pressure drops.

\begin{figure*}[t]
\centering
\includegraphics[width=\textwidth]{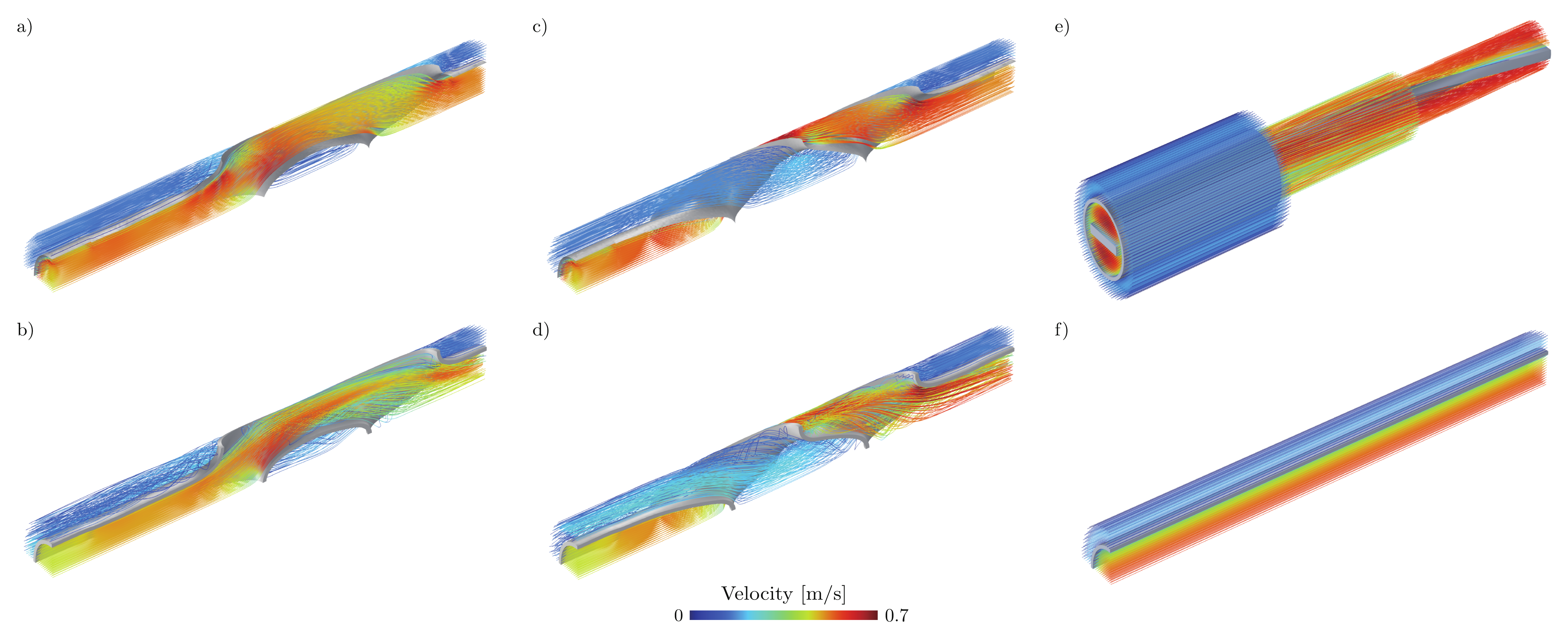}
\caption{Streamlines with velocity magnitude: a) the LF model of the optimized design at $\text{Re}=13{,}000$; b) the HF model of the optimized design at $\text{Re}=13{,}000$; c) the LF model of the optimized design at $\text{Re}=4{,}300$; d) the HF model of the optimized design at $\text{Re}=4{,}300$; e) the HF model of the twisted tape reference design; f) the HF model of the smooth pipe reference design.}
\label{fig:streamlines_vel}
\end{figure*}

\begin{figure*}[t]
\centering
\includegraphics[width=\textwidth]{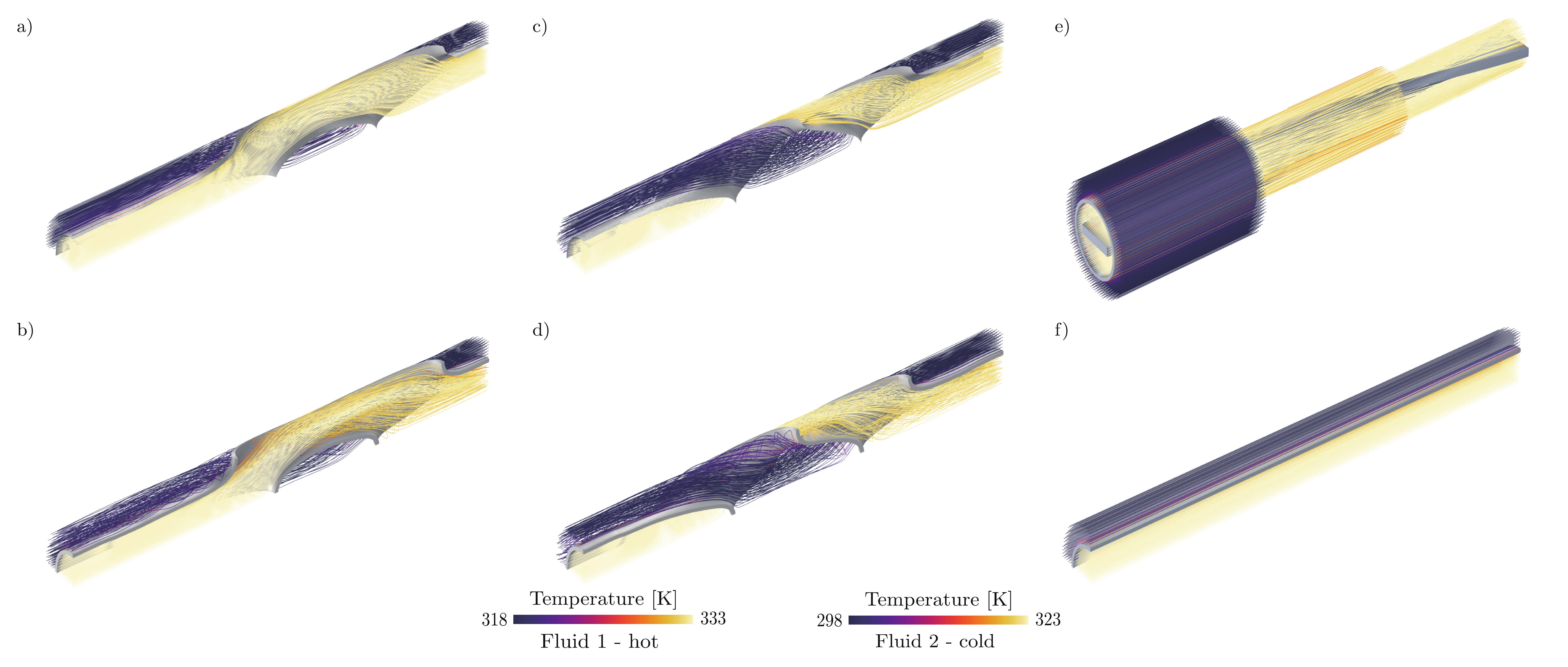}
\caption{Streamlines with temperature distributions: a) the LF model of the optimized design at $\text{Re}=13{,}000$; b) the HF model of the optimized design at $\text{Re}=13{,}000$; c) the LF model of the optimized design at $\text{Re}=4{,}300$; d) the HF model of the optimized design at $\text{Re}=4{,}300$; e) the HF model of the twisted tape reference design; f) the HF model of the smooth pipe reference design.}
\label{fig:streamlines_temp}
\end{figure*}

For better understanding of this performance evaluation, Figs.~\ref{fig:streamlines_vel} and \ref{fig:streamlines_temp} illustrate the streamlines with velocity magnitude and temperature distributions for the optimized designs same as those in Fig.~\ref{fig:PEC}, along with those for the corresponding LF models and reference designs, where all simulations are performed at $\text{Re}=13{,}000$ for fluid 1.
When comparing the velocity distribution of the twisted tape design shown in Fig.~\ref{fig:streamlines_vel}e with those of the smooth pipe design shown in Fig.~\ref{fig:streamlines_vel}f, the twisted tape induces moderate swirling flow in fluid 1 along the inserted tape, enhancing the convective heat transfer between fluid 1 and the solid wall.
However, this swirling flow is relatively weak due to the limited space in the pipe, resulting in only a moderate increase in overall heat transfer coefficient.
On the other hand, the optimized designs shown in Figs.~\ref{fig:streamlines_vel}b and \ref{fig:streamlines_vel}d exhibit increased surface area of the solid wall with complex flow patterns that induce strong swirling in both fluids, significantly enhancing the convective heat transfer.
Moreover, the designs optimized at $\text{Re}=4{,}300$ generate even stronger swirling flows and larger velocity values with more intricate solid wall structures, compared to those optimized at $\text{Re}=13{,}000$, as shown in Figs.~\ref{fig:streamlines_vel}b and \ref{fig:streamlines_vel}d, leading to higher overall heat transfer coefficients but higher pressure drops, as discussed above.
The LF model results in Figs.~\ref{fig:streamlines_vel}a and \ref{fig:streamlines_vel}c show the Darcy flow-based model can roughly capture the flow patterns and velocity magnitudes of the HF model, except for swirling flows induced near the solid wall surfaces and velocity gradients near the outer walls, as discussed in section~\ref{subsec:impl}.

Regarding the temperature distributions shown in Fig.~\ref{fig:streamlines_temp}, the smooth pipe design in Fig.~\ref{fig:streamlines_temp}f shows a relatively uniform temperature distribution due to the lack of flow disturbances.
The twisted tape design in Fig.~\ref{fig:streamlines_temp}e exhibits similar uniformity near the center of fluid 1 around the twisted tape, but shows cooled regions near the solid wall due to the enhanced convective heat transfer induced by the swirling flow.
This indicates that the twisted tape limits the mixing of fluid 1 across the pipe cross-section, resulting in less effective heat transfer enhancement.
The optimized designs in Figs.~\ref{fig:streamlines_temp}b and \ref{fig:streamlines_temp}d show significant temperature gradients near the solid wall regions, especially around the interfaces where the complex wall structures induce strong swirling flows.
These swirling flows promote effective mixing of fluid 1 and fluid 2 across the pipe cross-section, enhancing the convective heat transfer throughout the fluids.
The LF model results in Figs.~\ref{fig:streamlines_temp}a and \ref{fig:streamlines_temp}c show better agreement with the HF model results in terms of temperature distributions, compared to the velocity distributions, which is consistent with the discussion in section~\ref{subsec:impl}.

\subsubsection{Differences between LF and HF models}
\label{subsubsec:diff}

To quantitatively assess the differences between the LF and HF models for the optimized designs, we calculate the relative errors of pressure drops and temperature differences for both fluids, defined as follows:
\begin{align}
  \varepsilon_{\Delta p_i} & = \frac{|\Delta p_{i,\text{LF}} - \Delta p_{i,\text{HF}}|}{\Delta p_{i,\text{HF}}} \times 100\%, \\
  \varepsilon_{\Delta T_i} & = \frac{|\Delta T_{i,\text{LF}} - \Delta T_{i,\text{HF}}|}{\Delta T_{i,\text{HF}}} \times 100\%,
\end{align}
\noindent where the subscripts LF and HF denote the dimensional values obtained from the LF and HF models, respectively; $\Delta T_i = T_{\mathrm{out},i} - T_{\mathrm{in},i}$ is the temperature difference for fluid $i$.

Table~\ref{tab:error_stats} summarizes the mean and standard deviation of the errors for all optimized designs.
From Table~\ref{tab:error_stats}, the errors in pressure drops are generally larger than those in temperature differences.
This confirms the earlier discussion regarding the limitations of the Darcy flow-based LF model in accurately predicting pressure drops.
In contrast, the temperature differences are less affected by these simplifications, as heat transfer is more directly influenced by convective mixing and thermal conduction, which are better captured by the LF model.

\begin{figure*}[t]
\centering
\includegraphics[width=0.85\textwidth]{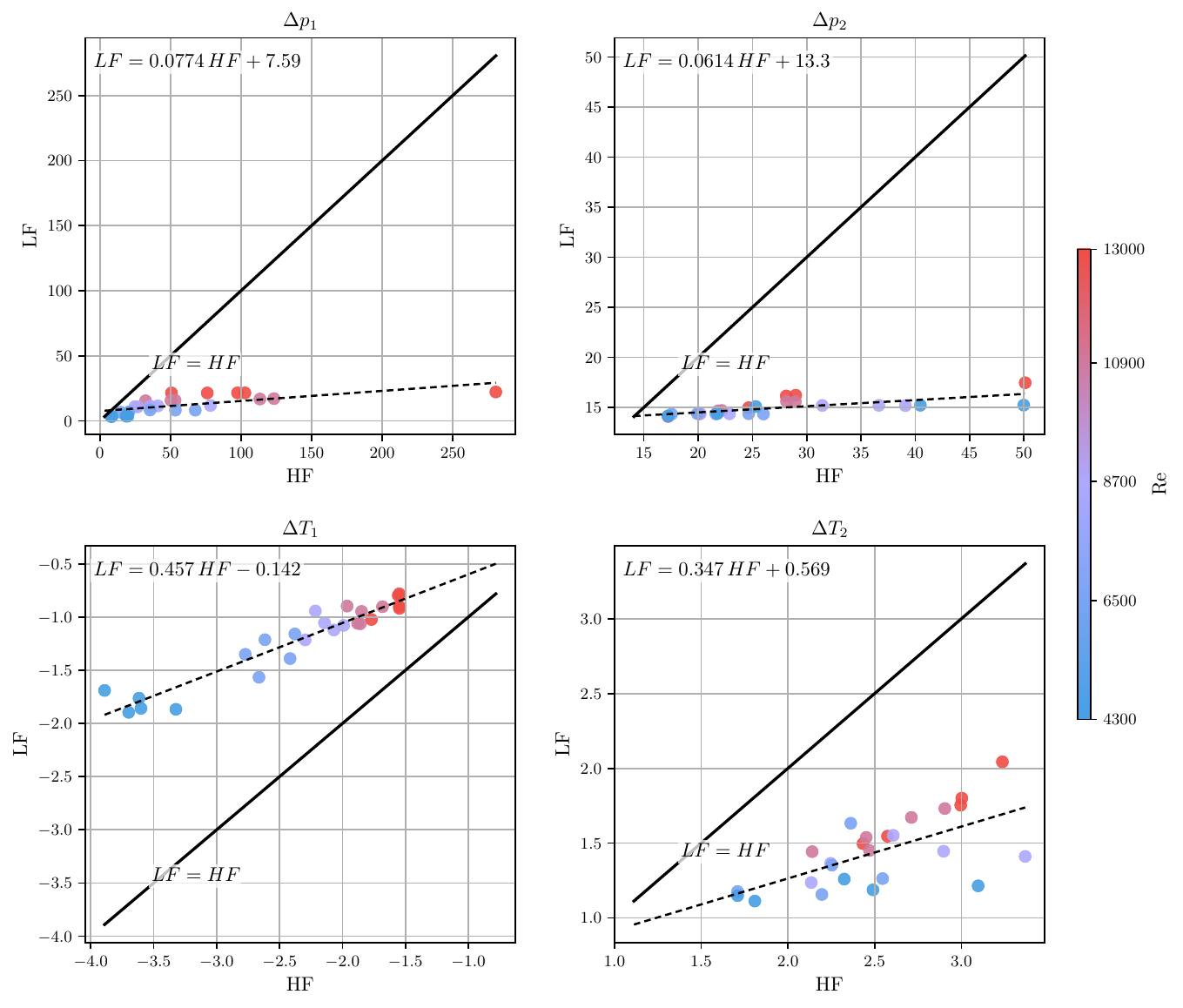}
\caption{Performance comparison between the optimized designs and reference design.}
\label{fig:parity_plot}
\end{figure*}

Fig.~\ref{fig:parity_plot} illustrates the parity plots of the pressure drops and temperature differences between the LF and HF models for all optimized designs, where the Reynolds number for fluid 1 is colored.
This parity plot indicates that when the plotted points are closer to the diagonal line, the LF model predicts the HF model results more accurately.
From Fig.~\ref{fig:parity_plot}, the LF model tends to underestimate the pressure drops for both fluids, as most points are located largely below the diagonal line in the pressure drop plots.
In addition, the errors in pressure drops for fluid 1 become larger as the Reynolds number increases, which is physically reasonable since higher Reynolds numbers lead to stronger turbulence and more complex flow patterns that are not well captured by the Darcy flow model.
The errors in pressure drops for fluid 2 are independent of the Reynolds number for fluid 1, as fluid 2 has a constant inlet velocity in all optimization cases.
In contrast, the temperature differences for both fluids show better agreement between the LF and HF models, with points closer to the diagonal line in the temperature difference plots, yet some underestimation is still observed.
This is likely due to the limitations of the LF model in capturing detailed convective mixing and thermal boundary layer effects, especially at higher Reynolds numbers where turbulence plays a significant role in heat transfer.
Moreover, the errors in temperature differences for fluid 1 tend to decrease with increasing Reynolds number, which is attributed to the increased velocity reducing the temperature decrease in fluid 1, thereby diminishing the relative impact of any discrepancies between the LF and HF models.
Contrally, the errors in temperature differences for fluid 2 tend to decrease with increasing Reynolds number for fluid 1, which is attributed to the trend of fluid 1 having more accurate temperature predictions at higher Reynolds numbers, thereby improving the prediction accuracy of the overall heat transfer for fluid 2.

\begin{table}[t]
\centering
\caption{Error statistics between LF and HF models for the optimized designs.}
{ \footnotesize
  \renewcommand{\arraystretch}{1}
  \begin{tabular}{lcccc}
\hline
 & $\varepsilon_{\Delta p_1}$ & $\varepsilon_{\Delta p_2}$ & $\varepsilon_{\Delta T_1}$ & $\varepsilon_{\Delta T_2}$ \\
\hline
Mean & 72.4\% & 40.9\% & 47.8\% & 41.9\% \\
Standard deviation & 11.9\% & 14.5\% & 4.8\% & 7.7\% \\
\hline
\end{tabular}}
\label{tab:error_stats}
\end{table}

\subsubsection{Influence of RANS modeling}
\label{subsubsec:RANS}

\begin{figure*}[h]
\centering
\includegraphics[width=\textwidth]{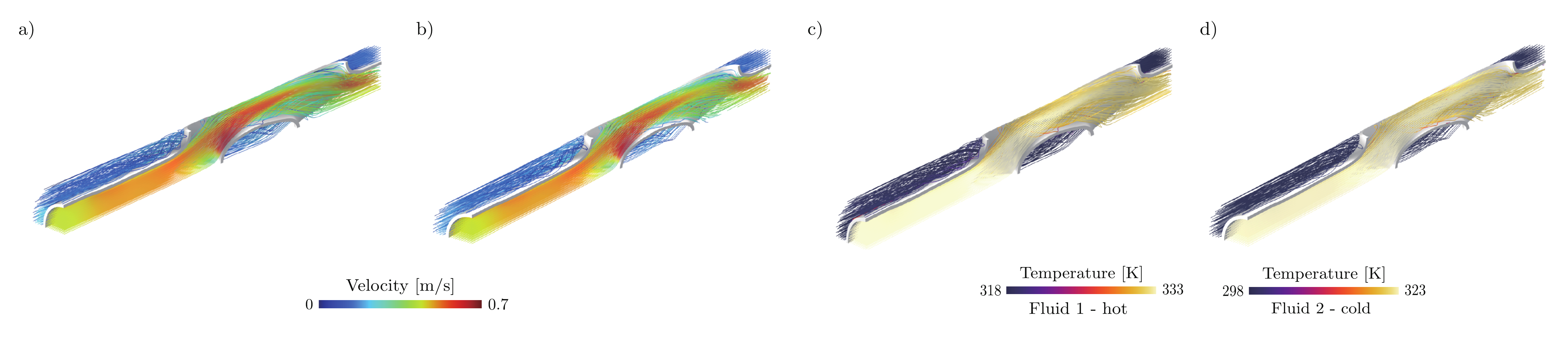}
\caption{RANS model comparison for the optimized design at $\text{Re}=13{,}000$ and $\alpha=1.0$: a) streamline with velocity magnitude using the $k$-$\omega$ SST model; b) streamline with velocity magnitude using the RSM; c) streamline with temperature using the $k$-$\omega$ SST model; d) streamline with temperature using the RSM.}
\label{fig:RANS_comparison}
\end{figure*}

To further investigate the influence of RANS modeling on the performance evaluation of the optimized designs, we compare the results obtained using three RANS turbulence models, i.e., the aforementioned k-$\omega$ model, the k-$\omega$ shear stress transport (SST) model, and the Reynolds stress model (RSM), on the optimized design at $\text{Re}=13{,}000$ with $\alpha=1.0$.
The k-$\omega$ SST model is a two-equation eddy viscosity model that combines the advantages of the k-$\omega$ model near the wall and the k-$\epsilon$ model in the free-shear region, often providing improved predictions for flows with adverse pressure gradients and separation~\cite{menter1994two}.
The RSM is a more high-fidelity turbulence model that solves transport equations for the Reynolds stresses, allowing for better predictions of anisotropic turbulence and complex flow phenomena, such as swirling flows and secondary flows~\cite{launder1975progress}.
Fig.~\ref{fig:RANS_comparison} illustrates the streamline with velocity magnitude and temperature distributions obtained from the k-$\omega$ SST model and RSM.
Compared to the k-$\omega$ model results shown in Figs.~\ref{fig:streamlines_vel}b and \ref{fig:streamlines_temp}b, the k-$\omega$ SST model results and the RSM results show closely similar flow patterns and temperature distributions, with some differences in the velocity magnitudes and temperature gradients, especially near the solid wall regions where turbulence effects are more pronounced.
To quantitatively compare the performance of these three RANS models, we calculate the pressure drops and temperature differences for both fluids using the k-$\omega$ SST model and RSM, and then compute the relative errors with respect to the k-$\omega$ model results, which are used as reference values, as shown in Table~\ref{tab:RANS_comparison}.
This table indicates that the k-$\omega$ SST model results are in good agreement with the k-$\omega$ model results, with errors less than $2.2$\% for all performance metrics, while the RSM results show slightly larger errors, especially for the fluid 1 that flows faster, with errors up to $2.9$\%.
This is reasonable since the k-$\omega$ SST model is still dependent on the k-$\omega$ model for turbulence modeling, while the RSM captures more complex turbulence phenomena that may not be fully represented in the k-$\omega$ model, leading to larger discrepancies in performance evaluation.
These results suggest that while the choice of RANS turbulence model can locally influence the performance evaluation of the optimized designs, the overall trends and conclusions regarding the performance improvements and trade-offs can be consistent across different turbulence models at least for the flow conditions considered in this study.
However, it is important to note that for flows with more complex turbulence phenomena, such as stronger swirling flows or flows with significant separation, the choice of turbulence model may have a more significant impact on the performance evaluation, and higher-fidelity models like RSM or even LES may be necessary for accurate predictions.
In that case, the designs optimized using the Darcy flow-based LF model may have larger discrepancies in performance evaluation when assessed with higher-fidelity turbulence models, which should be carefully considered in future studies.

\begin{table}[h]
\centering
\caption{Performance comparison of different RANS models.}
\label{tab:RANS_comparison}
{\footnotesize
\renewcommand{\arraystretch}{1.15}
\begin{tabular}{llcccc}
\hline
Model & Value/Error & $\Delta p_1$ [Pa] & $\Delta p_2$ [Pa] & $\Delta T_1$ [K] & $\Delta T_2$ [K] \\
\hline
k-$\omega$ & Value & 76.1 & 24.7 & -1.55 & 2.57\\
\hline
\multirow{2}{*}{k-$\omega$ SST}
& Value & 75.0 & 24.6 & -1.52 & 2.61\\
& Error & 1.4\% & 0.2\% & 2.2\% & 1.4\%\\
\hline
\multirow{2}{*}{RSM}
& Value & 74.2 & 24.3 & -1.6 & 2.6 \\
& Error & 2.6\% & 1.4\% & 2.9\% & 1.0\% \\
\hline
\end{tabular}
}
\end{table}

\section{Conclusions}
\label{sec:con}

In this study, we proposed a topology optimization method for designing two-fluid turbulent heat exchangers using a Darcy flow-based low-fidelity (LF) model to efficiently explore high-performance designs that enhance heat transfer while mitigating pressure drops.
The LF model was first calibrated against a high-fidelity (HF) model based on the Reynolds-averaged Navier-Stokes (RANS) equations to improve predictions of fluid flow and heat transfer characteristics.
Since the LF model, even after calibration, was expected to have discrepancies in performance predictions compared to the HF model, we adapted a multifidelity optimization approach, where the LF model was used for the optimization process and the HF model was employed for evaluating the optimized designs.
In the optimization process, to increase the likelihood of obtaining high-performance designs, we performed topology optimization for various inlet velocities and trade-off parameters to obtain diverse optimized designs.
The optimized designs were then evaluated using the HF model to assess their performance more accurately, where the overall heat transfer coefficients and pressure drops were compared with those of reference designs, including smooth pipes and twisted tape insertions.
Here are the key findings from this study:
\begin{itemize}
  \item The topology optimization using the Darcy flow-based LF model successfully generated various complex wall structures that enhance heat transfer while managing pressure drops evaluated by the RANS-based HF model, demonstrating the capability of the multifidelity optimization approach to explore high-performance designs efficiently.
  \item The Darcy flow-based LF model, after calibration, can reasonably predict the temperature distributions in turbulent two-fluid flows, although it tends to underestimate pressure drops due to the simplifications in modeling fluid flow.
  \item The optimized designs obtained from the LF model exhibited significant improvements in overall heat transfer coefficients, by up to $66.7$\%, compared to reference designs, while maintaining comparable or lower pressure drops, achieving up to $22$\% higher performance evaluation criteria (PEC) values.
  \item The performance evaluation revealed that high-performance designs optimized at higher Reynolds numbers tended to maintain high-performance across a wide range of Reynolds numbers, indicating their effectiveness in finding robust designs suitable for various operating conditions.
\end{itemize}

Despite these promising results, the proposed method still has some limitations, such as the reliance on manual selection of seeding parameters for the multifidelity optimization approach and the lack of consideration for practical design constraints like manufacturability and structural integrity.
Future work could focus on automating the multifidelity approach by a data-driven method~\cite{yaji2022data} to iteratively update the design based on the HF evaluations during the optimization process, reducing the dependence on manual parameter selection.
Moreover, adding more design constraints, such as manufacturability and structural integrity, could enhance the practical applicability of the optimized designs, followed by experimental validation of the proposed designs to confirm their performance in real-world applications.

\section*{Declaration of competing interest}
The authors declare that they have no known competing financial interests or personal relationships that could have appeared to influence the work reported in this paper.

\section*{Acknowledgements}
This work was supported by Innovative Future Space Transportation Systems Research and Development Program,
Japan Aerospace Exploration Agency (JAXA). 

\appendix
\section{Dimensionless variables}
\label{ap:dimensionless}
The variables in the Darcy flow-based model are non-dimensionalized as follows:
\begin{equation}
  \mb{x} = \frac{\mb{x}'}{X}, \quad \mb{u}_i=\frac{\mb{u}_i'}{U_i} , \quad p_i=\frac{p_i'}{\rho_i U_i^2}, \quad T= \frac{T'}{T_{\mathrm{ref}}}
\end{equation}
\noindent where the superscript $'$ denotes the dimensional variable; $X$ is the characteristic length; $U_i$ is the characteristic velocity; $T_{\mathrm{ref}}$ is the reference temperature.
In the present study, the characteristic length $X$ was set to the hydraulic diameter $D_\text{h}$, the characteristic velocity $U_i$ was set to the inlet velocity for fluid $i$ at the minimum inlet velocity among all optimization cases to be consistent; the reference temperature $T_{\mathrm{ref}}$ was set to the inlet temperature of fluid 2, i.e., $T_{\mathrm{in},2}$.

\section{Derivation of dimensionless energy equation for the Darcy flow-based model}
\label{ap:energy_equation}

The dimensional energy equation for the Darcy flow-based model can be expressed as follows:
\begin{equation}
  (\rho_1 {c_\text{p}}_1 \mb{u}_1' + \rho_2 {c_\text{p}}_2 \mb{u}_2') \cdot \nabla' T' - k_s C_k \nabla'^2 T' = 0,
\end{equation}
\noindent where the the first and second terms on the left-hand side represent the convective heat transfer for fluid 1 and fluid 2, respectively; the third term represents the conductive heat transfer in the solid wall; $\rho_i$ is the density; ${c_\text{p}}_i$ is the specific heat capacity; $C_k$ is the interpolated dimensionless thermal conductivity defined in Eq.~\eqref{eq:ck}.
By substituting the dimensionless variables defined in Appendix~\ref{ap:dimensionless} into the above equation, we can derive the dimensionless energy equation as follows:
\begin{equation}
  \left(\frac{\rho_1 {c_\text{p}}_1 U_1 X}{k_s} \mb{u}_1 + \frac{\rho_2 {c_\text{p}}_2 U_2 X}{k_s} \mb{u}_2\right) \cdot \nabla T - C_k \nabla^2 T = 0,
\end{equation}
\noindent where the dimensionless groups $\frac{\rho_i {c_\text{p}}_i U_i X}{k_s}$ can be recognized as the Péclet numbers $\text{Pe}_i$ for fluid $i$, leading to the final form of the dimensionless energy equation shown in Eq.~\eqref{eq:energy}.

\section{Derivation of the objective heat flux for the Darcy flow-based model}
\label{ap:heat_flux}

In the Darcy flow-based model, the objective heat flux for each fluid is defined as the total heat transfer from fluid $i$ to the solid wall region, normalized by the outlet area of fluid $i$.
Since the Darcy flow-based model does not provide detailed information about the local wall heat flux at the solid--fluid interface, we cannot directly evaluate the local wall heat flux.
Instead, the heat transfer is estimated from the net change in the sensible enthalpy carried by the fluid between the inlet and the outlet.

Assuming that the changes in kinetic energy, potential energy, viscous dissipation, pressure work, and axial heat conduction are negligible, the total heat transfer can be obtained from an energy balance over fluid domain $i$.
Therefore, the dimensional heat transfer rate is written as follows:
\begin{equation}
\dot{Q}_i = \int_{\Gamma_i^{\mathrm{out}}} \rho_i c_{p,i} (\mathbf{u}_i \cdot \mathbf{n}) \left( T - T_{\mathrm{in},i} \right) \, d\Gamma.
\end{equation}
The dimensional objective heat flux is then defined by normalizing $\dot{Q}_i$ with the outlet area:
\begin{equation}
  j_i^{h\prime} = \frac{1}{\int_{\Gamma_i^{\mathrm{out}}} d\Gamma'} \int_{\Gamma_i^{\mathrm{out}\prime}} \rho_i c_{p,i} (\mb{u}_i' \cdot \mb{n}) \left(T' - T_{\mathrm{in},i}'\right) \, d\Gamma'.
\end{equation}
Since the $d\Gamma' = X^{2} d\Gamma$, using the dimensionless variables introduced in~\ref{ap:dimensionless}, the above expression can be rewritten as follows:
\begin{equation}
  j_i^{h\prime} = \frac{\rho_i c_{p,i} U_i T_{\mathrm{ref}}}{\int_{\Gamma_i^{\mathrm{out}}} d\Gamma} \int_{\Gamma_i^{\mathrm{out}}} (\mb{u}_i \cdot \mb{n}) \left(T - T_{\mathrm{in},i}\right) \, d\Gamma,
\end{equation}
where $T_{\mathrm{in},i}=T_{\mathrm{in},i}'/T_{\mathrm{ref}}$ is the dimensionless inlet temperature.

To nondimensionalize the objective heat flux consistently with the energy equation in~\ref{ap:energy_equation}, the reference heat flux is chosen as $q_{\mathrm{ref}}=k_s T_{\mathrm{ref}}/X$.
Accordingly, the dimensionless objective heat flux is defined by the following equation:
\begin{equation}
  J_i^h = \frac{j_i^{h\prime}}{q_{\mathrm{ref}}} = \frac{X}{k_s T_{\mathrm{ref}}} j_i^{h\prime}.
\end{equation}
Substituting the dimensional expression into the above equation yields the following expression:
\begin{equation}
  J_i^h = \frac{\rho_i c_{p,i} U_i X}{k_s} \frac{1}{\int_{\Gamma_i^{\mathrm{out}}} d\Gamma} \int_{\Gamma_i^{\mathrm{out}}} (\mb{u}_i \cdot \mb{n}) \left(T - T_{\mathrm{in},i}\right) \, d\Gamma.
\end{equation}
Using the definition of the Péclet number $\mathrm{Pe}_i$, the dimensionless objective heat flux is finally written as Eq.~\eqref{eq:J}.

  \bibliographystyle{elsarticle-num} 
  \bibliography{reference}





\end{document}